\pdfoutput=1 
\documentclass[11pt]{article}
\usepackage[spanish, english]{babel}
\usepackage{amssymb}
\usepackage{amsmath}
\usepackage{stmaryrd}
\usepackage{subcaption} 
\usepackage{graphicx}
\usepackage{booktabs} 
\usepackage[font=small,skip=0pt]{caption}
\usepackage{lipsum}
\usepackage{epstopdf}
\usepackage{graphicx,epstopdf}
\def\theequation {\thesection.\arabic{equation}}
\makeatletter\@addtoreset {equation}{section}\makeatother

\def\ardlabel#1{\let\@currentlabel=\theequation\label{#1}}

\usepackage{graphicx}

\begin{document}

\title{\bf Linear guided modes and Whitham-Boussinesq model for variable topography}

\author{
R. M. Vargas-Maga\~na, A.A. Minzoni and
P. Panayotaros \\ 
{\small  Departamento de Matem\'aticas y Mec\'{a}nica IIMAS, Universidad Nacional Aut\'{o}noma de M\'{e}xico
} \\ 
{\small Apdo. Postal 20-726, 01000 M\'{e}xico D.F.,
M\'{e}xico } \\
{\small E-mail addresses: rmvargas@ciencias.unam.mx, panos@mym.iimas.unam.mx }\\
}
\date{}

\maketitle
%\line(1,0){450}

\begin{abstract}
In  this article we study two classical linear water wave problems,
i) normal modes of infinite straight
channels of bounded constant cross-section,
and ii) trapped longitudinal modes
in domains with unbounded constant cross-section.
Both problems can be stated using linearized free surface
potential flow theory, and our goal is to compare
known analytic solutions in the literature to numerical solutions
obtained using
an ad-hoc but simple
approximation of the non-local Dirichlet-Neumann operator
for linear waves proposed in
\cite{vargas2016whitham}.
To study normal modes
in channels with bounded cross-section we consider special symmetric triangular cross-sections, 
namely symmetric triangles with sides inclined at
$45^{\circ}$ and  $60^{\circ}$ to the vertical, and compare
modes obtained using the non-local Dirichlet-Neumann operator
to known semi-exact analytic expressions by 
Lamb \cite{lamb1932hydrodynamics}, Macdonald \cite{macdonald1893waves} ,
Greenhill \cite{greenhill1887wave},
Packham \cite{ packham1980small}, and Groves \cite{groves1994hamiltonian}.
These geometries have slopping beach boundaries that should in principle
limit the applicability of our approximate Dirichlet-Neumann operator. We nevertheless
see that the operator gives remarkably close results for even modes,
while for odd modes we have some discrepancies near the boundary.
For trapped longitudinal modes in domains
with an infinite cross-section we consider
a piecewise constant depth profile and
compare modes computed with the
nonlocal operator modes to known analytic solutions of
linearized shallow water theory by
 Miles \cite{miles1972wave}, Lin,  Juang and Tsay \cite{lin2001anomalous}, see also
\cite{mei2005theory}.
This is a problem of significant geophysical interest, and
the proposed model is shows to give quantitatively similar results
for the lowest trapped modes. 
\end{abstract}

\noindent {\bf Keywords:} linear water waves, variable topography, exact results, nonlocal shallow water wave models, 
transverse modes, continental shelves, triangular channels

\section{Introduction}

We study some problems on normal modes in linear water wave theory, in particular 
we compute numerically  
i) transverse normal modes in an infinite straight channel of bounded cross-section, 
and ii) trapped modes in domains with an unbounded cross-section. In both cases 
we consider special 
depth profiles for which we have known 
explicit or semi-explicit analytical solutions. 
The main goal of this paper
is to compare these solutions to numerical solutions obtained using 
a simple approximate nonlocal version of the water wave equations 
for variable depth that was    
recently proposed for dispersive shallow water waves \cite{vargas2016whitham}. 
In the case of transverse modes we compare modes of the approximate   
model of \cite{vargas2016whitham} 
to the known semi-analytic solutions for a channel with triangular cross-section
\cite{groves1994hamiltonian,packham1980small,lamb1932hydrodynamics,macdonald1893waves,greenhill1887wave}. 
In the study of trapped modes we compare the predictions of the 
model of \cite{vargas2016whitham}
to results obtained by the commonly used non-dispersive shallow water theory, 
applied to a model continental shelf geometry considered by several authors  
\cite{bonnet1990mathematical, kuznetsov2001spectrum, lin2001anomalous,mei2005theory, miles1972wave}.

The nonlocal linear system we use to compute normal modes is derived using the Hamiltonian 
formulation of the free surface potential flow \cite{zakharov1968stability}, see also 
\cite{radder1992explicit}, \cite{miles1977hamilton} and by approximating   
the (nonlocal) Dirichlet-Neumann operator for the Laplacian in the fluid domain appearing in the 
kinetic energy part of the Hamiltonian \cite{craig1994hamiltonian}.
Explicit expressions for the Dirichlet-Neumann operator in variable depth 
were derived by Craig, Guyenne, Nicholls and Sulem \cite{craig2005hamiltonian}, see also \cite{lannes2013water}.
Such expressions are generally complicated and  
can be used for numerical computations \cite{gouin2015development}, 
or to further simplify the equations of motion. 
\cite{vargas2016whitham} proposed an ad-hoc simplification of the variable depth Dirichlet-Neumann 
operator that leads to a simple variable depth generalization of Whitham-Boussinesq 
equations for shallow water theory. Nonlocal unidirectional and bidirectional 
shallow water models  
are of considerable current interest as 
nonlocal extensions of well-studied dispersive shallow water wave models such as the 
KdV and Boussinesq equations \cite{AcevesSanchez201380, moldabayev2014whitham}. We mention results 
on the existence of periodic traveling waves 
\cite{ehrnstrom2009traveling}, solitary waves
\cite{ehrnstrom2012existence}, wave breaking
\cite{naumkin1994nonlinear, constantin1998wave, hur2015breaking} 
and limiting Stokes waves \cite{ehrnstrom2016whitham}.  
The inclusion of variable depth effects is clearly of interest
in geophysical and coastal engineering applications, and raises additional questions
on the dynamics of these systems.

An immediate consequence of the Hamiltonian and Dirichlet-Neumann formulation  
of the water wave problem is that variable depth effects are already captured
at the level of the linear theory. For instance, the  
Dirichlet-Neumann operator can be expressed recursively in terms of the 
zero wave-amplitude  
Dirichlet-Neuman operator \cite{craig2005hamiltonian}. 
The study of linear normal modes is therefore   
a good test problem for comparing different approximations 
of the Dirichlet-Neumann operator for variable depth. 

The first part of the present work  
examines normal modes obtained by the approximate Dirichlet-Neumann operator
for special depth profiles that admit known semi-analytic normal mode solutions of the linearized 
water wave problem. These analytical results concern a few special depth profiles 
such as isosceles triangles 
with sides inclined at $45^{\circ}$ and  $60^{\circ}$ to the vertical,
see 
Greenhill  \cite{greenhill1887wave}, Macdonald  \cite{macdonald1893waves},
and the summary in Lamb's book, \cite{lamb1932hydrodynamics}. More recent studies are by 
Packham  \cite{ packham1980small} and Groves \cite{groves1994hamiltonian}.
A complete set of modes was also obtained for a semicircular channel
by Evans and Linton \cite{ evans1993sloshing}.
The construction yields odd and even normal modes that we then 
compare to odd and even eigenfunctions 
of suitable approximate Dirichlet-Neumann operator in a periodic domain (the period 
is the base of the triangle).
As we clarify in the next sections, the two problems are not equivalent because of the 
different assumptions at the intersection between the horizontal and sloping beach boundaries. 
(All known non-constant depth examples with analytic solutions concern domains with 
a sloping beach.)
Despite this problem, examining these examples we see that the two approaches
give comparable and often 
very close results for the normal mode shape, especially away from the beach. 
This is especially 
the case for higher even modes, where we see good agreement in the entire domain. 
Results for odd modes are close away from the 
beach, but have a marked discrepancy near the beach.  
We also see that approximating the triangular domain  
by a domain with a vertical wall (or a suitable periodic analogue)
yields modes that approach the triangular domain modes as the height of the wall
vanishes.
 
The second problem we study 
are trapped longitudinal modes in 3-D channels with a constant unbounded cross-section.  
The depth of the cross section is piecewise constant and takes two values,
with the smaller depth defined over a finite interval. 
Longitudinal modes are assumed to have a sinusoidal dependence in the longitudinal direction,
and 
trapped longitudinal modes are solutions that decay at infinity in the transverse horizontal 
direction.
The two-level step depth profile  
has been used by many authors Miles \cite{miles1972wave},  
Lin \textit{et al.} \cite{lin2001anomalous}, 
Mei \textit{et al.} \cite{mei2005theory}, to model localized waves that travel along continental 
shelves. These studies use a St. Venant-type linear shallow wave theory that leads 
to exact solutions and an algebraic determination of the number of trapped solutions with 
a given longitudinal speed.    
Trapped modes obtained using this approach are compared to numerical eigenfunctions  
of a higher dimensional analogue of the model Dirichlet-Neumann operator of \cite{vargas2016whitham}.  
We see a good qualitative agreement to the trapped modes computed exactly, but it remains an open
problem whether 
the two operators predict the same number of trapped states.    

The organization of the paper is as follows. 
In \emph{Section 2} we formulate the linear water wave problem
and present the nonlocal operators used to approximate the linear system. 
In \textit{Section 3} we consider triangular depth geometries, review some of the  
known semi-analytic normal mode solutions, and compare them to  
to numerical eigenmodes computed using the approximate nonlocal operator.
In \textit{Section 4}  we compute the semi-exact longitudinal mode 
solutions of the shallow water theory in a simple geometry used in the literature 
to model two continental 
shelves, and compare with results obtained using the nonlocal model.

\section[Formulation of the problem]{Formulation of the problem and approximate 
\\ Dirichlet-Neumann operators}\label{formulation}

\begin{figure}
    \centering
    \begin{subfigure}[b]{0.45\linewidth}        %% or \columnwidth
        \centering
        \includegraphics[scale=0.3]{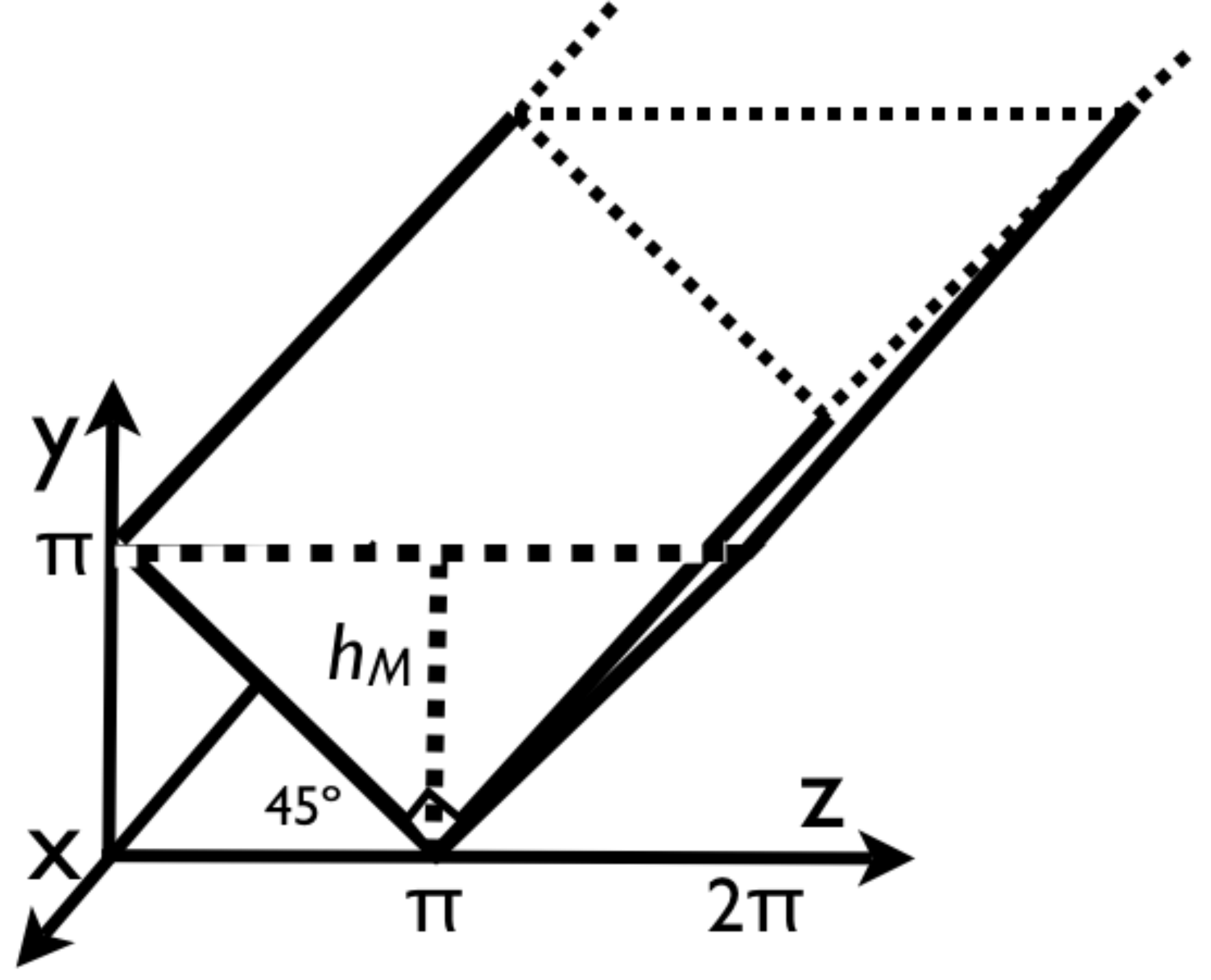}
        \caption{$\beta_{45}$}
      \label{fig:T45}
    \end{subfigure}
    \begin{subfigure}[b]{0.45\linewidth}        %% or \columnwidth
        \centering
        \includegraphics[scale=0.25]{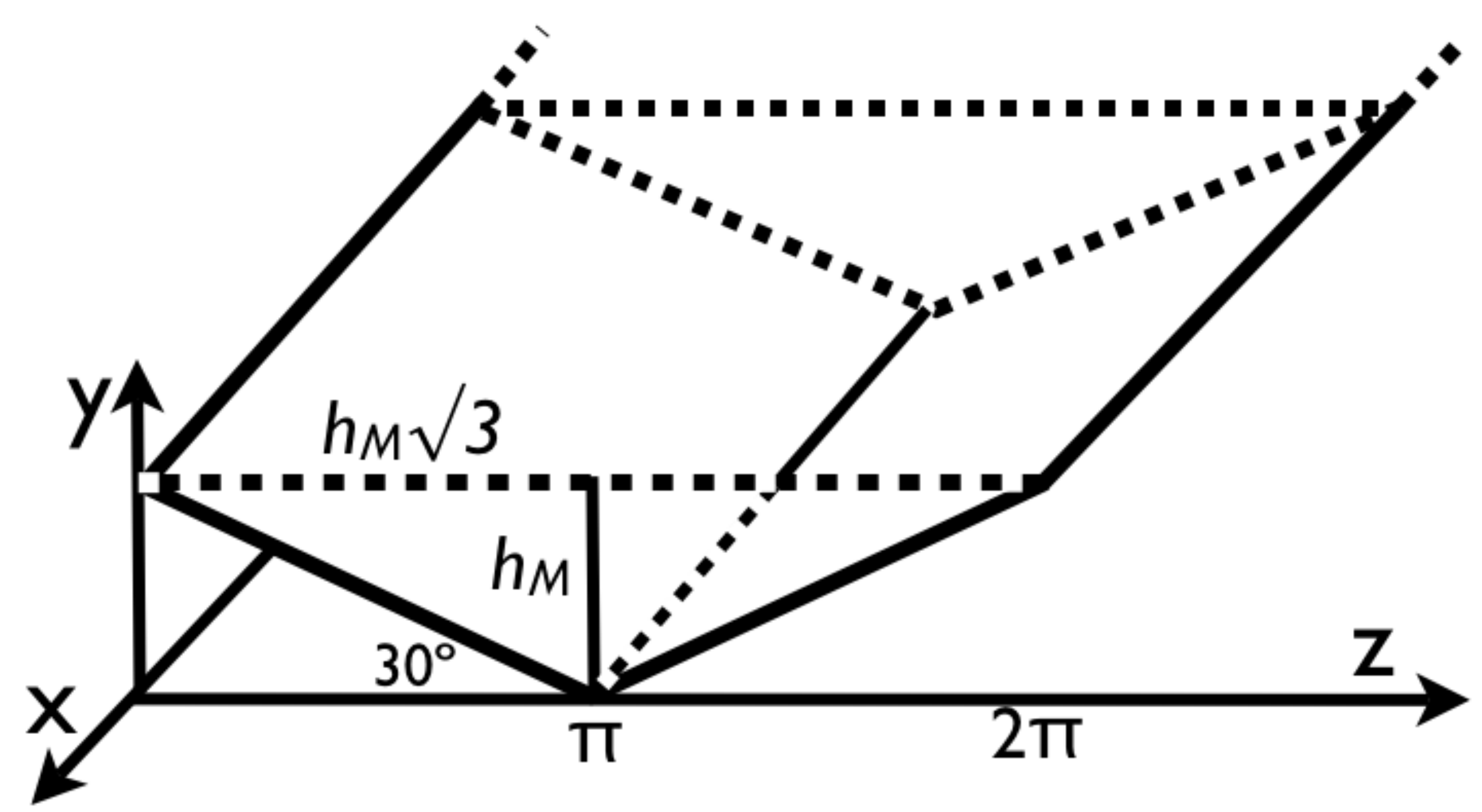}
        \caption{$\beta_{30}(z)$}
        \label{fig:T30}
    \end{subfigure}
    \caption{Schematics of straight channels with triangular cross-sections $\Omega$ and coordinate system}
    \label{fig:Tchannels}
\captionsetup{width=0.4\textwidth}
\centering
		\includegraphics[scale=0.3]{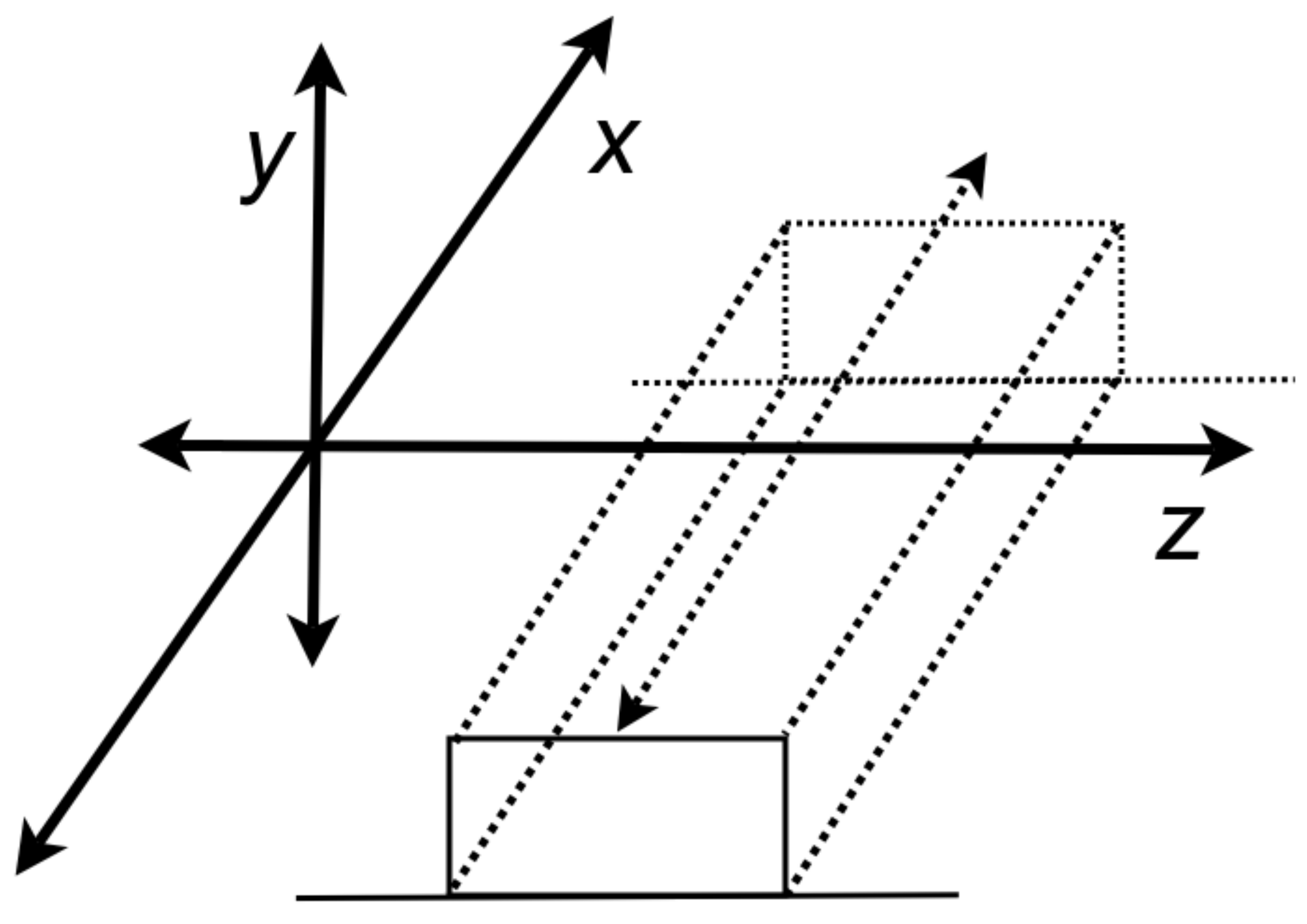}
\caption{Schematics of submerged continental shelf $\beta_{h_i}(z)$ and coordinate system.} 	
    \end{figure}

The problems considered in this paper come
from the classical linear theory of water waves \cite{lamb1932hydrodynamics}.
To describe the equations we use Cartesian coordinates 
denoted by $(x,y, z)$, where $y$ is directed vertically upwards,
$x$  is measured longitudinally along the channel and $z$
is measured across the channel or across the submarine ridge, 
see e.g. \figurename{s \ref{fig:T45}, \ref{fig:T30}, \ref{ContinentalShelf}}.

We define the fluid domain  as
$\mathcal{D}= \mathbb{R} \times \Omega,$ where $\Omega$ is the cross section,
and we distinguish bounded and unbounded cross sections $\Omega = \Omega_B$, $\Omega_U$
respectively, with 
\begin{equation}\label{omega-bounded}
 \Omega_B = \lbrace [z,y]: z \in [0,b], y \in  [h_m+\beta(z), h_M], 
\end{equation}
\begin{equation}\label{omega-unbounded}
 \Omega_U = \lbrace [z,y]: z \in \mathbb{R}, y \in  [h_m+\beta(z), h_M].
\end{equation}
The heights $y = h_m$ and $y = h_M$ describe the minimum and maximum elevations respectively, with 
$h_m < h_M$, and $h_m + \beta(z) \leq h_M$ for all $z$.

We will assume $\partial \mathcal{D}= \mathbb{R} \times \partial \Omega =\Gamma_{L} \cup  \Gamma_{F} \cup \Gamma_{B}$,  
with $\Gamma_L$ representing the lateral wall, $\Gamma_F$ representing the free surface and $\Gamma_B$ the bottom. 

In the case $\Omega= \Omega_B$ we have
\begin{equation}
\Gamma_F= \lbrace (x,h_M,z): x\in \mathbb{R}, 0\leq z \leq b \rbrace,
\end{equation}
 \begin{equation}
\Gamma_B= \lbrace (x, h_{m}+\beta(z), z): x\in \mathbb{R}, 0\leq z \leq b  \rbrace,
\end{equation}
 \begin{equation}
\Gamma_L= \lbrace (x,y,0): x\in \mathbb{R}, y \in [h_m+ \beta(0), h_M]\rbrace \cup 
\lbrace (x,y,b): x\in \mathbb{R}, y \in [h_m+ \beta(b), h_M]\rbrace.
\end{equation}
In this article are primarily interested in domains with $h_m + b(z) = h_M$ at $z=0$ and $b$. 
Then $\Gamma_L= \emptyset$.

%We are thus assuming that $\partial \mathcal{D}$ has no lateral walls.  

In the case $\Omega= \Omega_U$ we are interested in domains with $\beta(z) \rightarrow 0$ as 
$z \rightarrow \pm \infty$.

To state the problem we introduce a velocity potential $\phi(x,y,z,t)$
and look for solutions of  Laplace's equation
\begin{equation}\label{Lap}
\phi_{xx}+ \phi_{yy} +\phi_{zz}=0 \text{ in } \mathcal{D,}
\end{equation}
with 
\begin{equation}\label{Ebc}
 \left\lbrace \begin{array}{l} 
\phi_t + g\eta= 0 \text{ on } \Gamma_F,  \\   
\eta_t= \varphi_y \text{ on } \Gamma_F ,\\  
\frac{\partial \phi}{\partial {\hat n}}=0 \text{ on } \Gamma_B \cup \Gamma_L, \end{array} \right.
\end{equation}
see \cite{whitham2011linear}. Equations 
\eqref{Lap}, \eqref{Ebc}
 are the linearized Euler equations 
for free surface potential flow. 

\label{ContinentalShelf}

We consider solutions of the following two forms:

\noindent A. 
\begin{equation}\label{TransverseMode}
\phi (x,y,z,t)=\psi(y,z)\cos(\omega t), 
\end{equation} 
referred to as \textit{transverse  modes}  see \cite{packham1980small}, \cite{groves1994hamiltonian}.
 
\noindent B. 
\begin{equation}\label{LongMode}
\phi(x,y,z,t)=\psi(y,z)\cos(\kappa x-\omega t), 
\end{equation} 
referred to as \textit{longitudinal  modes}, see \cite{packham1980small} \cite{kuznetsov2002linear}.

We reformulate the problem of finding solutions of the above form in terms of the 
Dirichlet-Neumann operators.

We first consider transverse modes.
Let the fluid domain consist of a straight channel and consider the case $\Omega = \Omega_{B}.$ 
Consider $f: [0,b] \longrightarrow \mathbb{R}$, and define 
the Dirichlet-Neumann operator 
$G_{0}$ by 
\begin{equation}\label{OpGo}
(G_0 [f])(z) =g\frac{\partial \psi(z,y)}{\partial y} |_{y = 0 },
\end{equation}
where $\psi: \Omega \rightarrow \mathbb{R}$ satisfies
 \begin{equation}\label{EqUPplugging}
 \left\lbrace \begin{array}{l} 
\Delta\psi= 0\text{ } \text{ at }\Omega, \\  
%\frac{\partial \psi}{\partial y}= \frac{ \omega^2}{g}\psi(z,0)   %\text{  at } \Omega \cap \Gamma_F ,\\ 
\psi(z,y) = f  \text{ at } \Omega \cap \Gamma_F , \\ 
\frac{\partial \psi}{\partial {\hat n}}=0 \text{ } \text{ at }\Omega \cap (\Gamma_B \cup \Gamma_L) \\ 
%\psi \in H^1(\Omega), \psi \neq 0,\\
 \end{array} \right.
\end{equation}
By \eqref{Lap}-\eqref{Ebc}, 
the problem of finding 
transverse mode solutions \eqref{TransverseMode} 
is then equivalent to solving  
 \begin{equation}
\label{DN-eigenfunctions}
(G_0 [f])(z) =\omega^2 f.
\end{equation}

Boundary conditions for $f$ must be specified consistently 
with \eqref{EqUPplugging}.
Note that the original formulation \eqref{Lap}-\eqref{Ebc} does not
require any boundary conditions on $f$, unless 
$\Gamma_L$ is nonempty, in which case 
we add Neumann conditions $f_z = 0$ at $z =0$, $b$.

The operator $G_0$ will be approximated by 
\begin{equation} \label{operadorAG}
\mathcal{A}_{G_0}[f]= \textit{Sym}(D \tanh (h(z)D))[f], 
\end{equation}
using the notation 
%and definitions 
%of \cite{vargas2016whitham} for the $b-$periodic case, 
\begin{equation}\label{PDOazD}
  a(z,D)[f](z)  
=\frac{1}{b} \sum \limits_{k=-\infty}^{\infty}  a(z,k)   \hat{f}_k  e^{ik\frac{2 \pi}{b}z},
\end{equation}
\begin{equation}
\hat{f}_k= \frac{1}{b} \int_{-{\frac{b}{2}}}^{\frac{b}{2}} f(z) e^{-ik\frac{2\pi}{b}z}dz
\end{equation}

$\mathcal{A}_{G_0}$ is an ad-hoc approximation of $G_0$
obtained from the constant depth expression by making $h$ variable, and symmetrizing.  
$\mathcal{A}_{G_0}$ satisfies some basic properties of the exact $G_0$, e.g. symmetry, 
correct high-wavelength asymptotics, see \cite{vargas2016whitham},
and also has a simple form.   
As it becomes clearer in the next section the $b-$periodic  
boundary conditions here are not entirely appropriate.  
Also, the domains we consider in the next section are symmetric in $z$ so that 
the eigenfunctions of $ \mathcal{A}_{G_0}$ are either even or odd, satisfying 
Neumann and Dirichlet boundary conditions respectively at $z = 0$, $b$.   
This suggests that in the presence of lateral boundaries we should also consider only the even modes as physical.

We now consider longitudinal modes for the 
case $\Omega_{B}.$ Consider $f(z)=\psi(0,z)$, 
with $z \in [0,b],$ 
and define the 
modified Dirichlet-Neumann operator $G_\kappa$ by 
 \begin{equation}\label{operadorgorro}
 (G_{\kappa} [f])(z) = g \frac{\partial \psi(z,y)}{\partial y} |_{y = 0 }
\end{equation}
where $\psi : \Omega \longrightarrow \mathbb{R}$ satisfies
\begin{equation}\label{EqTrans}
 \left\lbrace \begin{array}{l} 
\Delta\psi=\kappa^2 \psi \text{ } \text{ in }\Omega, \\  
%\frac{\partial \psi}{\partial y}= \frac{\omega^2}{g}\psi(0,z)   %\text{ at } \Omega \cap \Gamma_F , \\ 
\psi(z,y)=f \text{ in } \Omega \cap \Gamma_F , \\
\frac{\partial \psi}{\partial {\hat n}}=0 \text{ } \text{ at }\Omega \cap (\Gamma_B \cup \Gamma_L).\\ 
 \end{array} \right.
\end{equation}
The problem of finding longitudinal mode
solutions \eqref{LongMode} 
of \eqref{Lap})-\eqref{Ebc}
is equivalent to the spectral problem  
 \begin{equation}\label{modified-DN-eigenfunctions}
(G_{\kappa} [f])(z) =\omega^2 f.
\end{equation}
Boundary conditions on $f$
are as in the case of transverse modes.

The operator $G_{k}$ of \eqref{operadorgorro} will be approximated by the operator
\begin{equation}\label{Aproxgorro}
 {\mathcal{A}}_{G_{\kappa}}(\beta)= \textit{Sym}[\sqrt{\kappa^2 + D^2} \tanh(h(z) \sqrt{\kappa^2 + D^2})], \text{ } \text{}
\end{equation}
with $ h(z)=h_0 - \beta(z) $, $h_0 = h_M - h_m >  0$ and $D= -i\partial_z$ see \eqref{PDOazD}. 
The operator ${\mathcal{A}}_{G_{\kappa}}(\beta)$ is 
obtained heuristically in the same way as ${\mathcal{A}}_{G_{\kappa}}$,
generalizing the case of constant depth.

Longitudinal modes for the 
case $\Omega_U$ lead to \eqref{EqTrans} with $ f(z)=\psi(0,z)$, $z \in (-\infty,\infty)$, 
$\Gamma_L = \emptyset$.
Solutions of \eqref{modified-DN-eigenfunctions} that decay at infinity will be referred to it, 
see \cite{ursell1952edge}, \cite{bonnet1990mathematical},  \cite{kuznetsov2002linear}
for examples in semi-infinite geometry.

${\mathcal{A}}_{G_{\kappa}}(\beta)$ can be also obtained by 
considering the three dimensional analogue of the approximate 
Dirichlet-Neumann operator of the previous chapter of the form 
\begin{equation}\label{Aproxgorro-3D}
 {\mathcal{A}}_{G_0}(\beta)= \textit{Sym} [\sqrt{- \Delta} \tanh(h(z) \sqrt{-\Delta})], 
\end{equation}
where $\Delta $ is the Laplacian in $(x,z)$, 
and $ h(z)=-h_0 + \beta(z) $, $h_m = - h_0 < 0$. We note that the depth varies only in the $z$ direction. 
Generalized eigenfunctions of the operator (\ref{Aproxgorro}) of the form $f(z) e^{ \pm i \kappa x}$
lead to the spectral problem for the operator ${\mathcal{A}}_{G_{\kappa}}(\beta)$ of (\ref{Aproxgorro}).

In the next section we describe some known 
semi-analytic solutions of the form \eqref{TransverseMode}, \eqref{LongMode}.
obtained for special domains with finite cross-section 
$\Omega$ and $\Gamma_L = \emptyset$, i.e. slopping beach geometries.
These solutions are constructed starting with a multiparameter 
family of harmonic functions $\phi_\mu$ of the form 
\eqref{TransverseMode}, or \eqref{LongMode},
defined on the plane and satisfying the rigid wall boundary condition on a 
set ${\tilde \Gamma}_B$ that includes 
$\Gamma_B$ and 
is the boundary of a domain 
$\tilde{\mathcal{D}}$ that includes $\mathcal{D}$.  
Then we require that the $\phi_\mu$ also satisfy the 
first two equations of \eqref{Ebc} on the free surface $\Gamma_F$. 
This requirement leads to algebraic equations 
that restrict the allowed values of the parameter $\mu$ 
to a discrete set and also determine the 
frequencies. This construction does not assume any boundary conditions 
for the potential at the free surface.

\section{Transverse and longitudinal modes in triangular cross-sections}\label{Sec3p3}
  
Transverse and longitudinal modes can be calculated explicitly only
for special geometries of the channel cross-sections. 
In this section we compare some exact results for triangular channels by 
Lamb \cite{lamb1932hydrodynamics} \textit{Art. 261},  Macdonald \cite{macdonald1893waves}, 
Greenhill \cite{greenhill1887wave}, Packham \cite{ packham1980small}, and Groves \cite{groves1994hamiltonian}, 
to results obtained using the non-local operator \eqref{Aproxgorro}.

\subsection{Transverse modes for triangular cross-section: $45^{\circ}$ case}

\begin{figure}%[h!]
\captionsetup{width=0.71\textwidth}
\centering
\includegraphics[scale=0.32]{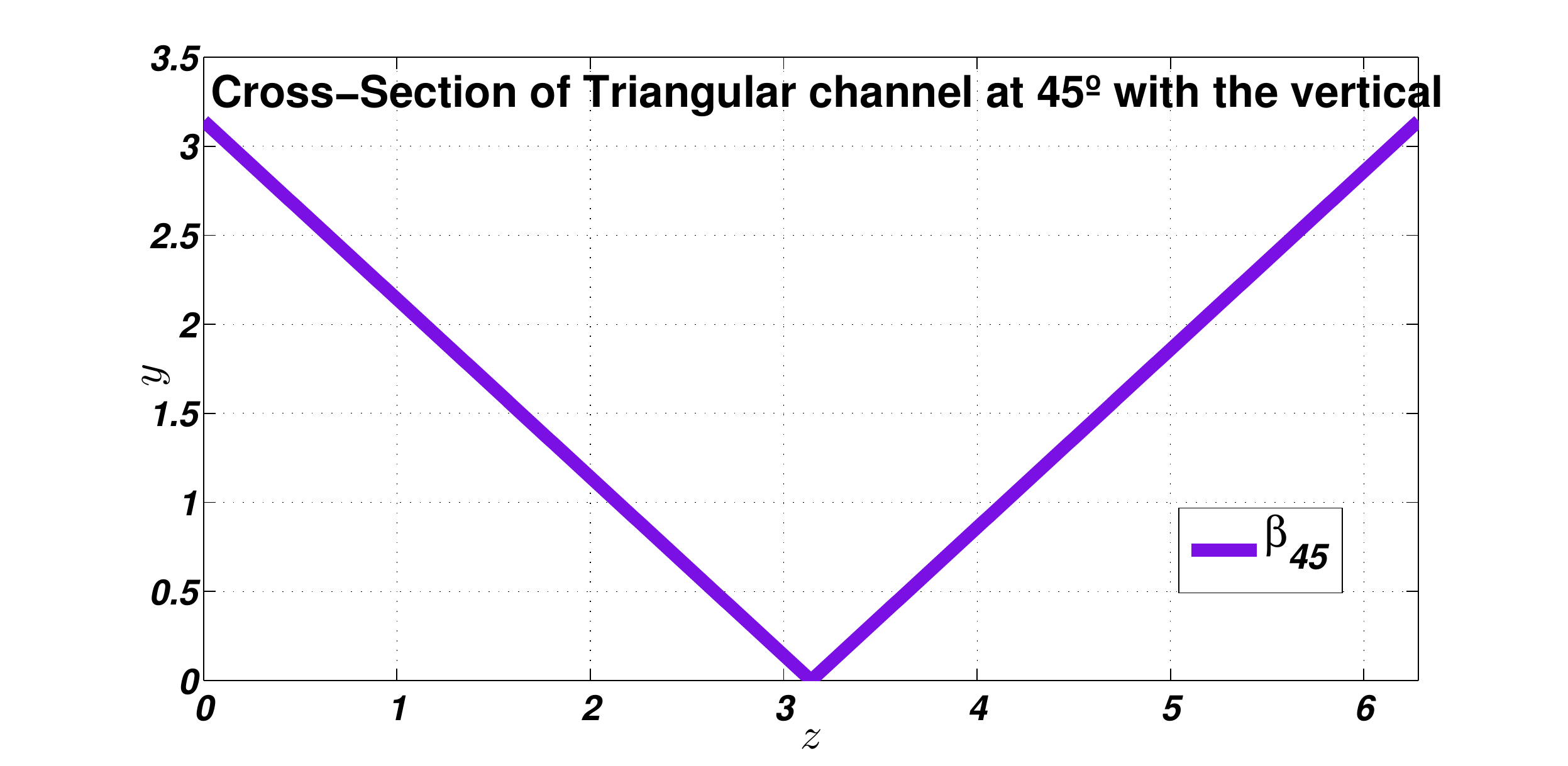}
\caption{Triangular Cross-section of a straight channel see equation \protect\eqref{beta45}.} \label{Figbeta45}	
\captionsetup{width=0.71\textwidth}
%\centering
\end{figure}

The first geometry we consider corresponds 
to a uniform straight channel with triangular cross-section  
with a  semi-vertical angle $45^{\circ}$, see \figurename{ \ref{Figbeta45}}. 
The cross-section $\Omega= \Omega_{B}$ is as in  \eqref{omega-bounded} and the bottom is 
at $y = \beta_{45}(z)$. The minimum and maximum heights of the the fluid domain are 
$h_m = 0$ and $h_{M}=\pi$ respectively 
The  channel width is  $b=2\pi$, and
\begin{equation} \label{beta45}
 \beta_{45}(z) =\left\lbrace \begin{array}{l}   
-z + \pi \text{ } \text{ in }  0 \leq z < \pi \\
z -\pi \text{ } \text{ in }  \pi \leq z \leq 2\pi \\ 
 \end{array} \right. , \text{ } z \in [0, 2\pi]. 
\end{equation}

\begin{figure}%[h!]
    \centering
    \begin{subfigure}[b]{0.48\linewidth}        %% or \columnwidth
        \centering
        \includegraphics[scale=0.27]{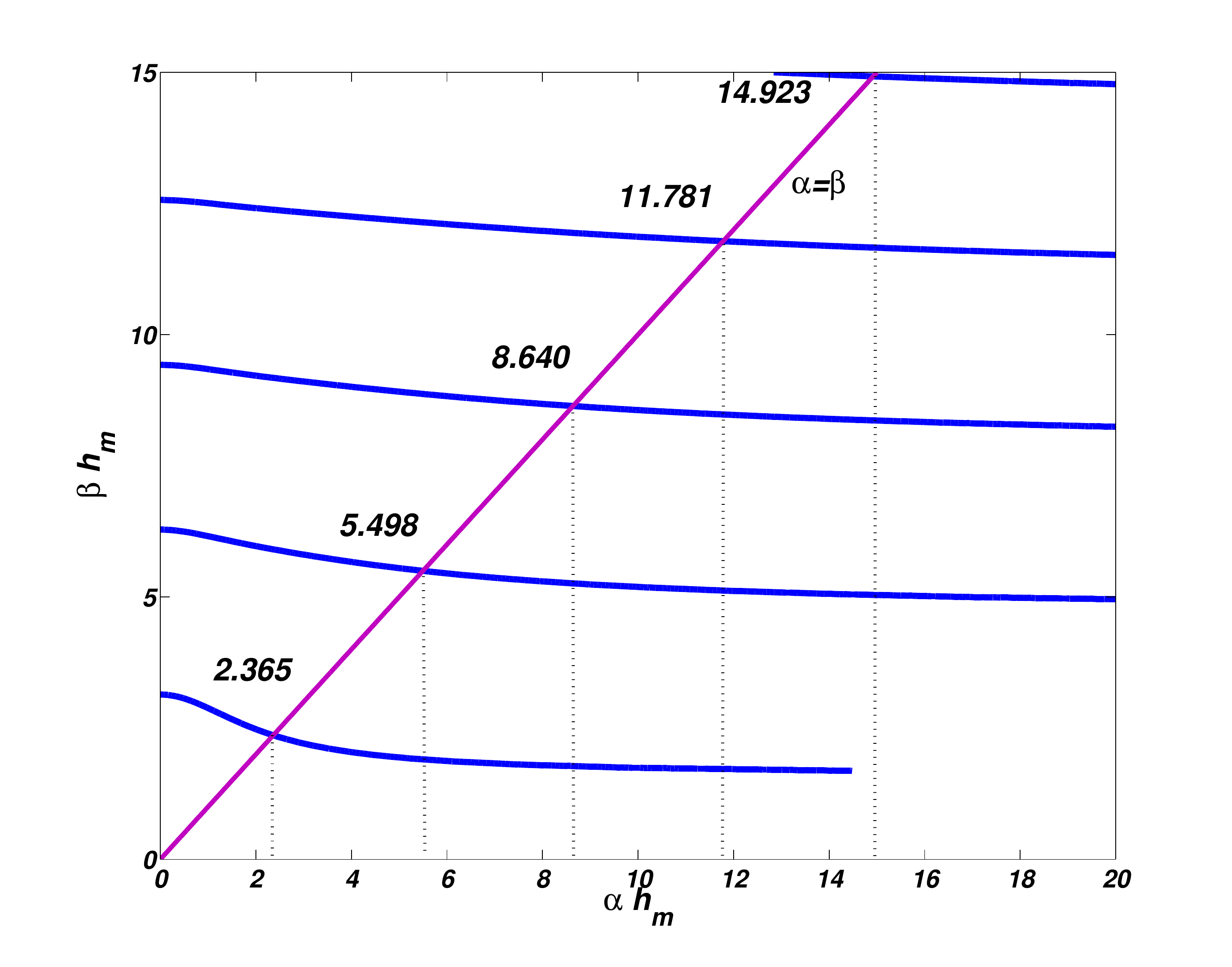}
        \caption{Even modes}
        \label{fig:Even45}
    \end{subfigure}
    \begin{subfigure}[b]{0.48\linewidth}        %% or \columnwidth
        \centering
        \includegraphics[scale=0.32]{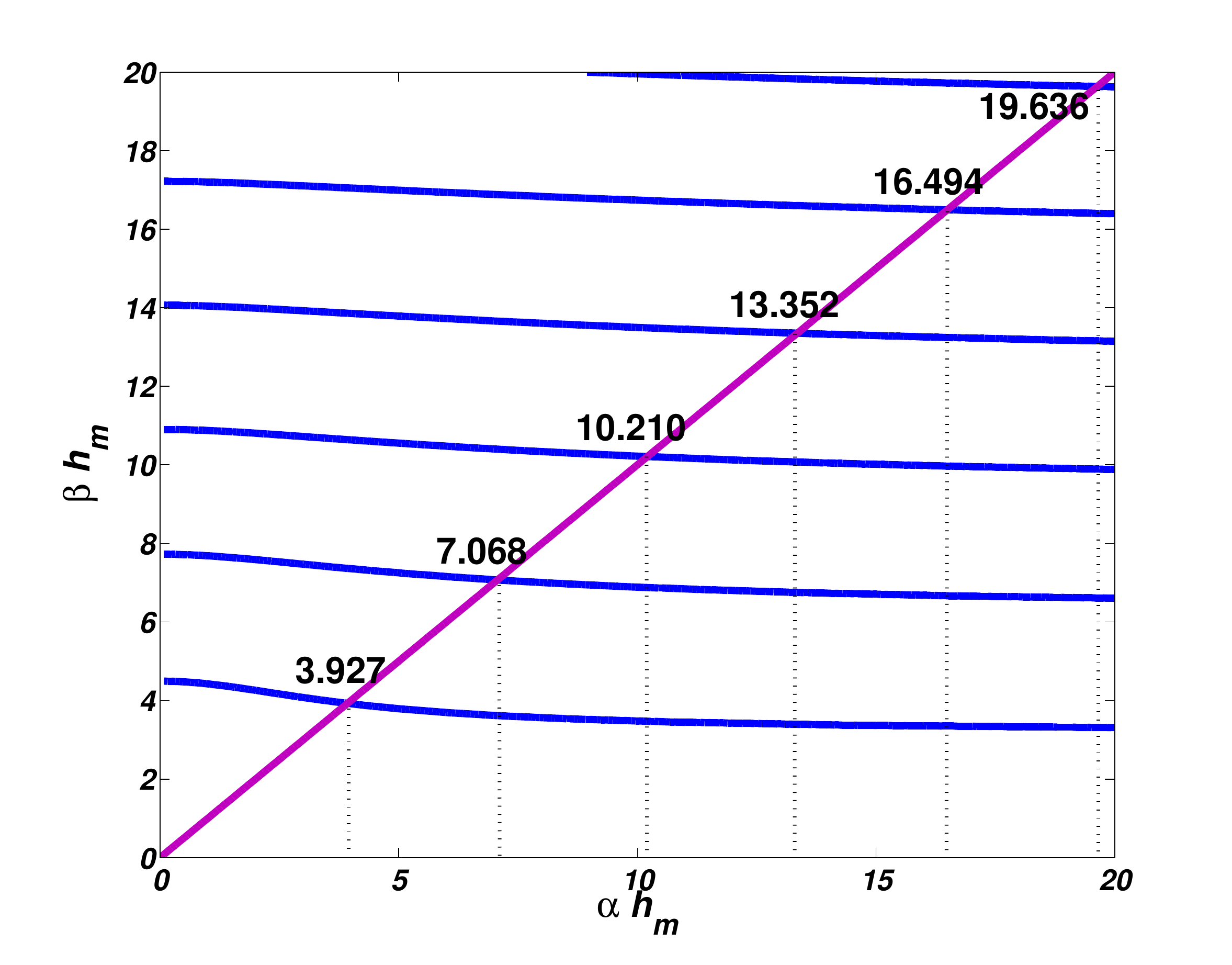}
        \caption{Odd modes}
        \label{fig:Odd45}
    \end{subfigure}
    \caption{(a)  Intersection of curves \protect\eqref{kzero357} and \protect\eqref{curva357}. (b) 
   Intersection  of curves: \protect\eqref{kzero246} and \protect\eqref{curva246}. }
    \label{alphabeta45}
\end{figure}

Normal modes for this channel were obtained by Kirchhoff, see Lamb \cite{lamb1932hydrodynamics}, 
\textit{Art. 261}, and include symmetric and antisymmetric modes. We review their results:

The  symmetric transverse modes, see \eqref{TransverseMode}, are given by the potential
\begin{equation}\label{SymModes}
\psi = A[\cosh(\alpha z)\cos(\beta y) + \cos(\beta z)\cosh(\alpha y)]\cos(\omega t).  
\end{equation}
It can be checked that $\frac{\partial \phi}{\partial y} \mid_{y=0}$. Also,
$\psi$ is symmetric with respect to the $z = 0$ axis, is  
harmonic in the quarter plane $y \geq |z|$, and 
satisfies the rigid wall boundary condition 
\begin{equation}
\frac{\partial \phi}{\partial \hat{n}}=0  \quad \text{at} \quad y= |z|.
\end{equation}
To impose the boundary condition at the free surface $y = h_M$, we   
use the first two equations of \eqref{Ebc} to obtain   
\begin{equation}\label{SymModesSurface}
\omega^2  \psi = \frac{1}{g} \frac{\partial \psi}{\partial y}   \quad \text{at} \quad y= h_M. 
\end{equation}\label{kzero246}
Combining with \eqref{SymModes} we have the conditions 
\begin{equation}
\alpha^2 - \beta^2 =0, 
\end{equation}
and 
\begin{equation}\label{omega246}
\omega^2 \cosh(\alpha h_M)= g\alpha \sinh(\alpha h_M), 
\quad 
\omega^2\cos(\beta h_M)=-g\beta \sin(\beta h_M)=0, 
\end{equation}
or 
\begin{equation}\label{curva246}
\alpha h_M \tanh (\alpha h_M) + \beta h_M \tan (\beta h_M)=0.
\end{equation}
The values of $\alpha$, $\beta$ are determined by the intersections of the curves 
\eqref{kzero246} and \eqref{curva246}, see 
\figurename{ \ref{fig:Even45}}. There is an infinite number of solutions,
$h_M \alpha_j$, $j= 0,2,4,\ldots$,  with $\alpha_{j}< \alpha_{j'}$ if $j<j'$. The 
corresponding frequencies $\omega_j$ are obtained by \eqref{omega246}. 
The first values of $h_{M}\alpha_i, \omega_i$ are shown in Table 1. % \ref{tab:table3p145}.
These values are obtained from the intersections of curves \eqref{kzero357} and \eqref{curva357} and of
curves \eqref{kzero246} and \eqref{curva246} shown in
\figurename{ \ref{fig:Even45}} and \figurename{ \ref{fig:Odd45}} respectively. 

To obtain 
the antisymmetric modes we use the potential
\begin{equation}\label{AsymModes}
\phi= B(\sinh(\alpha z) \sin(\beta y)+ \sin(\beta z) \sinh(\alpha y))\cos(\omega t).
\end{equation}
We check that $\phi $
satisfies the rigid wall 
boundary conditions at $y = |z|$ and is harmonic in the quadrant $y \geq |z|$.
We also check that $\frac{\partial \phi}{\partial y} \mid_{y=0}$ is antisymmetric with respect to the $z = 0$ axis. 
Imposing the free surface boundary conditions \eqref{Ebc} to \eqref{TransverseMode} we obtain  
\eqref{SymModesSurface}. Then \eqref{AsymModes}
leads to the conditions  
\begin{equation}\label{kzero357}
\alpha^2 - \beta^2=0
\end{equation}
and
\begin{equation}\label{omega357}
\omega^2 \sinh(\alpha h_M)= g\alpha \cosh(\alpha h_M),   
 \quad \omega^2 \sin(\beta h_M)= g\beta \cos(\beta h_M),
\end{equation}
or 
\begin{equation}\label{curva357}
\alpha h \coth(\alpha h_M)= \beta h_M \cot(\beta h_M).
\end{equation}
The values of $\alpha$, $\beta$ are determined by the intersections of 
the curves \eqref{kzero357} and \eqref{curva357}, see \figurename{ \ref{fig:Odd45}}. % \ref{tab:table3p145}.
There is an infinite number of solutions 
$h_M\alpha_{j}$, $j={1,3,5...}$ with $\alpha_{j}< \alpha_{j'}$ if $j<j'$. 
The corresponding frequencies $\omega_{j}$ are given by \eqref{omega357}, see Table 1. 

By \eqref{Ebc}
the free surface corresponding to the above symmetric and antisymmetric
modes is computed by 
\begin{equation}\label{freeS}
\eta(z)= \frac{1}{g} \frac{\partial \phi}{\partial y} \mid _{y=h_M}.
\end{equation}

\begin{table}%[h!]
%\label{tab:table3p1}  
  \centering
  \captionsetup{width=0.71\textwidth}
  \caption{Frequencies of modes for channel of \figurename{ \ref{Figbeta45}}}
  \begin{tabular}{cccccccc}
   \toprule
$\text{ } $ &$i=0$ & $i=1$& $i=2$ & $i=3$ & $i=4$& $i=5$ & $\cdots$ \\
   \midrule
    $\alpha_i h_M$ & 2.365 & 3.927 & 5.498 & 10.210 & 11.781& 16.494 & $\cdots$\\
    $\omega_i$ &4.8624   & 7.8261  &4.1413 & 5.6445  &6.0622 & 7.1684  & $\cdots$\\
  \bottomrule
  \end{tabular}
  %\caption*{}
\end{table}\label{tab:table3p145}

In \figurename{ \ref{T45Modes01234}} and \figurename{ \ref{T45AsymModes01234}} 
we compare the surface amplitude of 
the exact symmetric and antisymmetric modes found above to the 
surface amplitudes obtained by computing numerically 
the eigenfunctions of the approximate Dirichlet-Neumann operator 
$ {\mathcal{A}}_{G_{0}}(\beta)$ of \eqref{operadorAG}
with $2 \pi-$periodic boundary conditions. The operator is discretized 
spectrally.  
Given a computed eigenfunction $f$ of $ {\mathcal{A}}_{G_{0}}(\beta)$ 
we obtain the surface amplitude $\eta$
by $\eta = \frac{1}{g}\mathcal{A}_{G_{k}}(\beta)$. This is analogous to \eqref{freeS}. 

\figurename{ \ref{T45Modes01234}} suggests
good quantitative agreement for the even modes. 
For the odd modes we see that the wave amplitudes differ at the boundary representing the sloping beach. 
In particular the odd eigenvectors of $ {\mathcal{A}}_{G_{\kappa}}(\beta)$ 
have nodes at $z = 0$, $b = 2\pi$, while the exact odd modes do not. 
The procedure for obtaining 
the exact modes does not require any conditions on the value
of the potential at the intersection of the free boundary and the rigid wall. 
Also, the free surface is described by a value of the $y-$coordinate, and this allows 
us to define $\eta(z)$ for all real $z$, in particular 
we determine the fluid domain by computing the intersection 
of the graph of $\eta$ with the rigid wall.
This leads to a more realistic motion of the surface at the sloping beach, 
although this does not imply that the exact solutions are physical either, since the 
boundary conditions at the free surface are not exact. 
In contrast,
the odd modes of $ {\mathcal{A}}_{G_{\kappa}}(\beta)$ correspond to 
pinned boundary conditions that are not expected to be physical.    
\begin{figure}%[h!]
\captionsetup{width=0.75\textwidth}
\centering
		\includegraphics[scale=0.6]{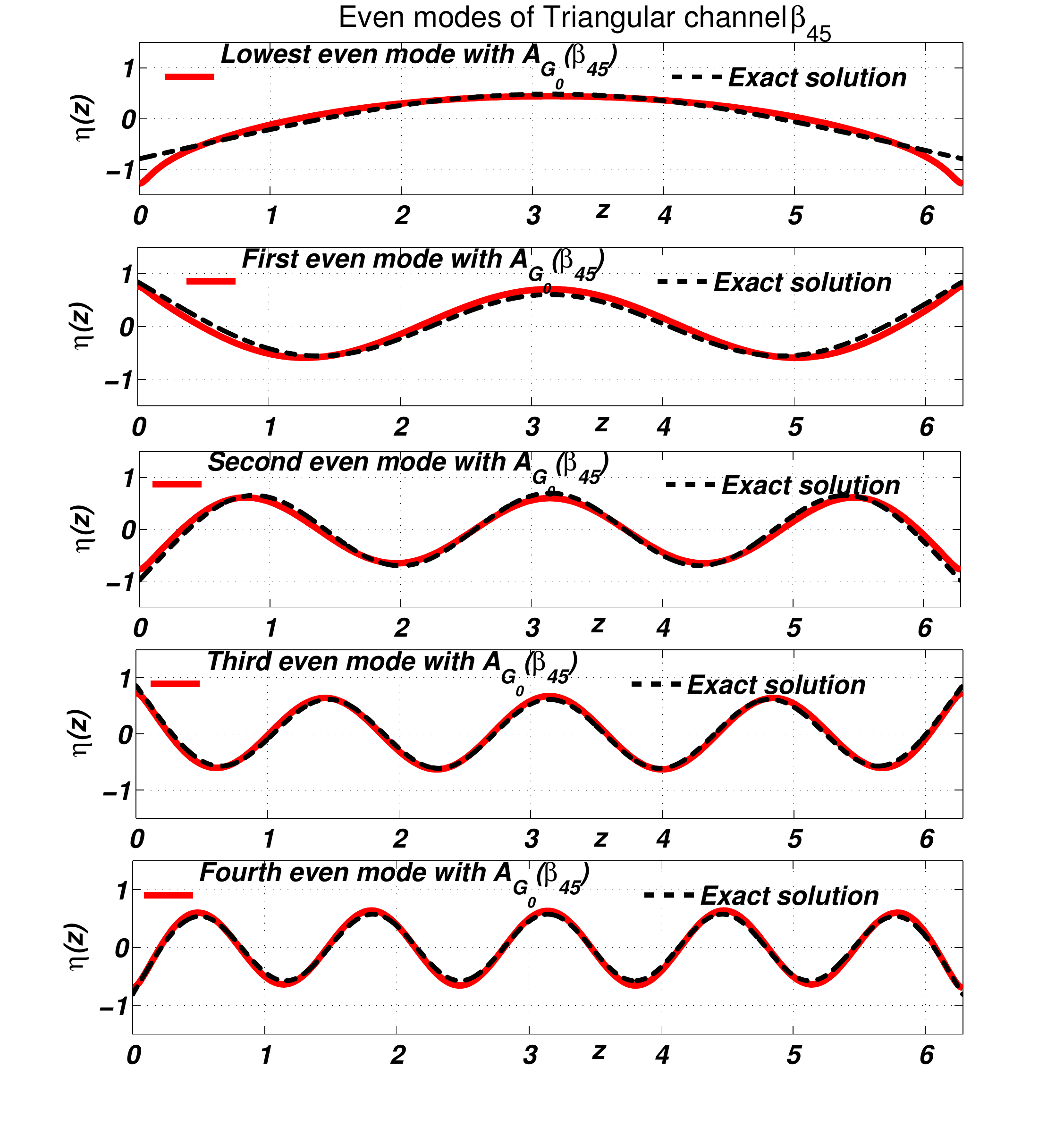}
\caption{Symmetric transverse modes of a channel with triangular cross 
section illustrated in \figurename{ \protect\ref{Figbeta45}}. 
Red (solid) lines: operator $\mathcal{A}_{G_0}$, dashed lines: 
exact solutions from \protect\eqref{SymModes}  and the values in Table 1.} %\ref{tab:table3p145}.}
\end{figure}\label{T45Modes01234}
\begin{figure}%[h!]
\captionsetup{width=0.75\textwidth}
\centering
		\includegraphics[scale=0.6]{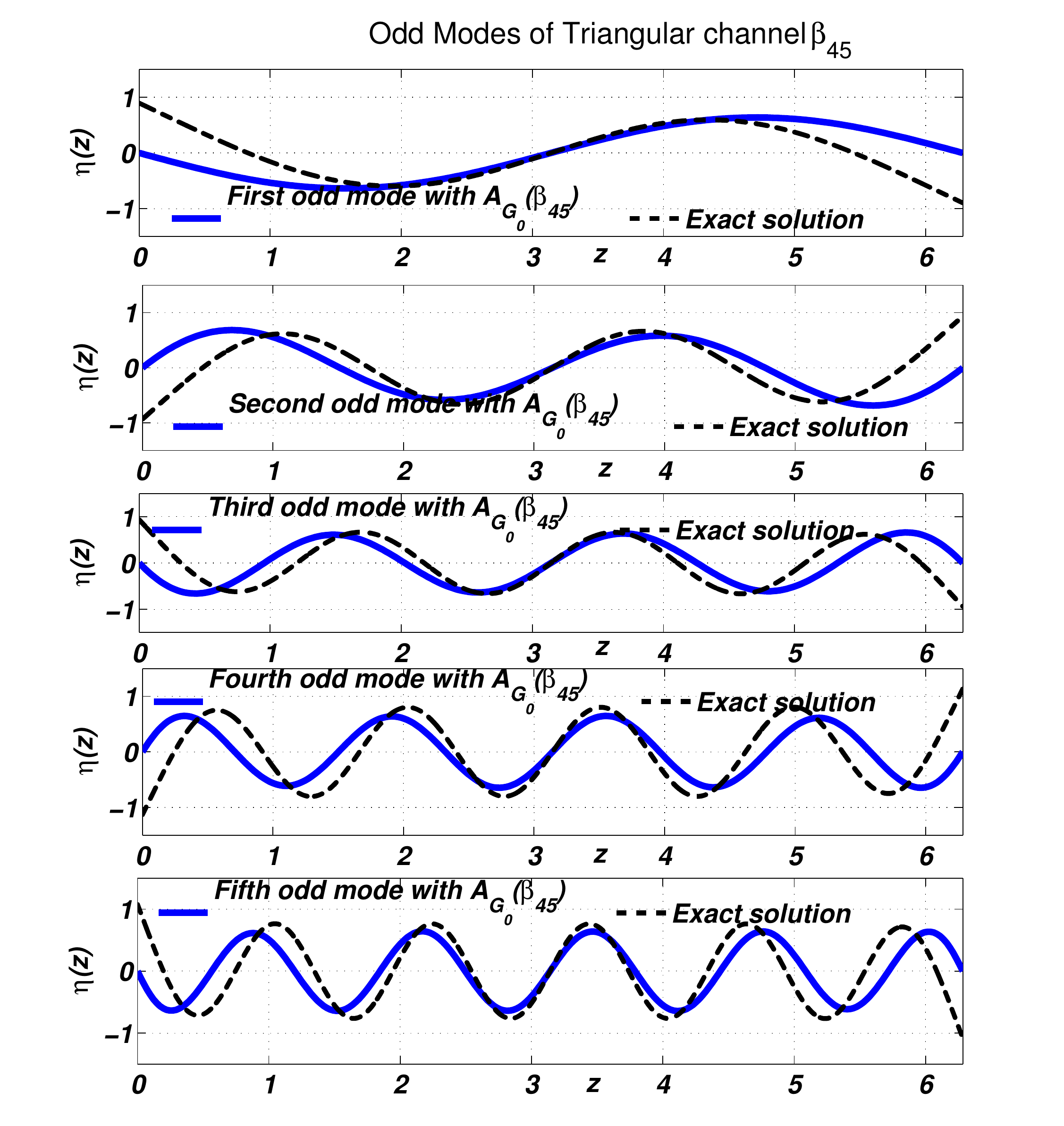}
\caption{Antisymmetric transverse modes of a channel with triangular cross section
illustrated in \figurename{ \protect\ref{Figbeta45}}.
Blue (solid) lines: operator $\mathcal{A}_{G_0}$, dashed lines:
exact solutions using \protect\eqref{AsymModes} and the values in Table 1.
%\ref{tab:table3p145}.
}
\end{figure}	\label{T45AsymModes01234}

\subsection{Longitudinal modes for triangular cross-sections: $60^{\circ}$ case}

\begin{figure}
\includegraphics[scale=0.42]{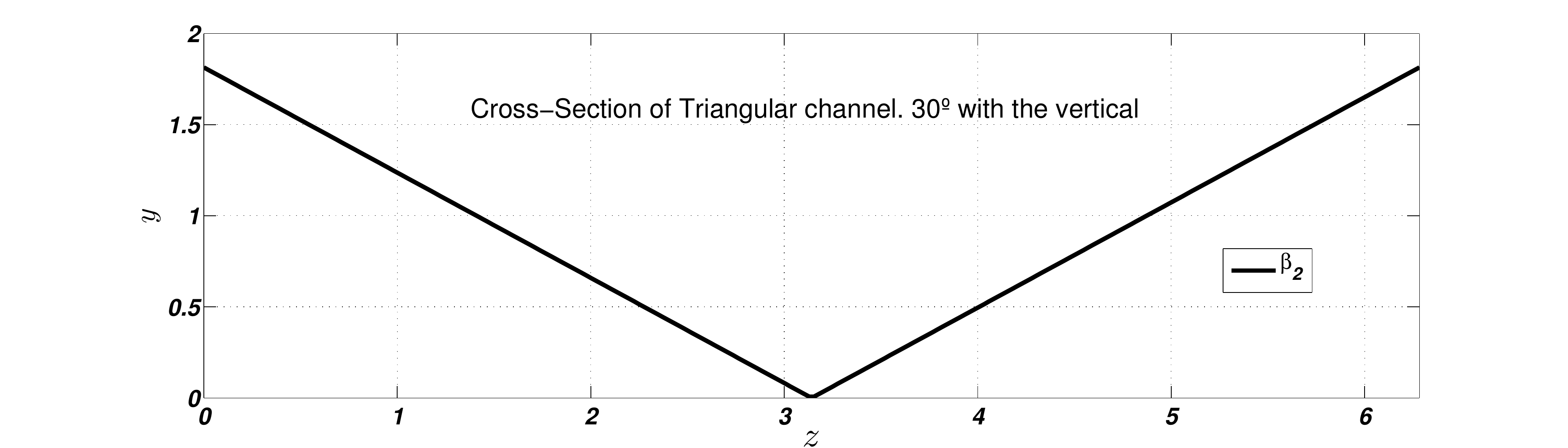}
\caption{Triangular cross-section of a straight channel, see equation \protect\eqref{beta30}.} 	
\captionsetup{width=0.71\textwidth}
\label{Figtrian30}
\end{figure}

A second geometry with exact longitudinal modes was 
considered by Macdonald \cite{macdonald1893waves}, Packham, \cite{ packham1980small}, see also 
Lamb \cite{lamb1932hydrodynamics}, \textit{Art. 261}. This geometry 
corresponds to a uniform straight channel with triangular cross-section  
with a  semi vertical angle $60^{\circ}$, as illustrated in \figurename{ \ref{Figtrian30}}. In this case we will examine longitudinal modes. 

We consider the cross-section $\Omega= \Omega_{B}$, as in \eqref{omega-bounded} and 
the bottom  $\beta_{30}(z)$. 
The  channel width is  $b=2\pi$ and the maximum and minimum heights of the fluid domain are
$h_{M}=\frac{\pi}{\sqrt{3}}$ and $h_m = 0$ respectively. 
The cross-section profile is given by 
\begin{equation} \label{beta30}
 \beta_{30}(z) =\left\lbrace \begin{array}{l}   
\frac{-1}{\sqrt{3}}z + \frac{\pi}{\sqrt{3}} \text{ } \text{ in }  0 \leq z < \pi \\
\frac{1}{\sqrt{3}}z -\frac{\pi}{\sqrt{3}} \text{ } \text{ in }  \pi \leq z \leq 2\pi \\  
 \end{array} \right. , \text{ } z \in [0, 2\pi].
\end{equation}

The exact  solutions for symmetric modes, see Packham \cite{ packham1980small}, 
and Groves \cite{groves1994hamiltonian} are the following. 

The $0-$mode is described by a velocity potential $\phi$ of the form \eqref{LongMode} with
\begin{eqnarray}\label{lowestmode}
\psi(z,y) &=& A[\cosh (\kappa(y-h_M))+ 
\frac{\omega^2 \kappa^2}{g\kappa}\sinh(\kappa(y-h_{M}))\\ \nonumber
& &+ 2\cosh(\frac{\sqrt{3}\kappa (z-\pi)}{2})\\ \nonumber
& &\times \lbrace \cosh(\kappa (\frac{y}{2}+ h_M)) - 
\frac{\omega^2 \kappa^2}{g\kappa}\sinh(\kappa (\frac{y}{2}+ h_M))\rbrace ],
\end{eqnarray}
and 
\begin{equation}
\label{lowestmodefreq}
\omega^2= \frac{3g}{4\kappa}\coth(\frac{3\kappa h_M}{2})\lbrace1
+ (1-\frac{8}{9} \tanh^2(\frac{3\kappa h_M}{2}))^{\frac{1}{2}}\rbrace.
\end{equation}

The remaining symmetric modes $2,4,6,8,...$ are
 described by a velocity potential $\phi$ of the form \eqref{LongMode} with
\begin{eqnarray}\label{sym30}
\psi &=& A[\lbrace \cosh(\alpha(y-h_M)) +\frac{\omega^2 \kappa^2}{g\kappa} \sinh(\alpha(y-h_M)) \rbrace \cos(\beta(z-\pi)) \\ \nonumber
& &+ 2\cosh(\frac{\sqrt{3}\alpha (z-\pi)}{2}) \cos(\frac{\sqrt{3}\beta y}{2}) \cos(\frac{\beta (z-\pi)}{2})\\\nonumber
& &\times\lbrace \cosh(\alpha(\frac{y}{2}+h_M)) - \frac{\omega^2 \kappa^2}{g\kappa} \sinh(\alpha (\frac{y}{2} + h_M)) \rbrace \\\nonumber
& & -2 \sinh(\frac{\sqrt{3}\alpha (z-\pi)}{2}) \sin(\frac{\sqrt{3}\beta y}{2} \sin(\frac{\beta (z-\pi)}{2}))\\\nonumber
& & \times \lbrace \sinh(\alpha(\frac{y}{2}+h_M)) - \frac{\omega^2 \kappa^2}{g\kappa}\cosh(\alpha(\frac{y}{2}+h_M))   \rbrace],
\end{eqnarray}
\begin{equation}\label{omegasym30}
\omega^2 =\frac{g\alpha}{\kappa^2}\left[
\frac{\frac{\beta}{\alpha}\sqrt{3}(\cosh(3\alpha h_M)- \cos(\sqrt{3}\beta h_M))}{\frac{\beta}{\alpha}\sqrt{3}\sinh(3\alpha h_M)-3 \sin(\sqrt{3}\beta h_M)}
\right].
\end{equation}
The above potentials are harmonic and satisfy the rigid wall boundary conditions.  
The first two equations of motion \eqref{Ebc} lead to   
\begin{equation}\label{alfabetakappa}
\alpha^2- \beta^2= \kappa^2,
\end{equation}
and
\begin{eqnarray}\label{relalfabetakappa}
& &\left( \frac{\beta}{\alpha}\right) ^2 \cosh(3\alpha h_M)\cos(\sqrt{3}\beta h_M)\\ \nonumber
&-&\frac{1}{4}\left( \frac{\beta}{\alpha}\right)  \sqrt{3} \left\lbrace  1-\left( \frac{\beta}{\alpha}\right) ^2 \right\rbrace \sinh(3\alpha h_M)\sin(\sqrt{3}\beta h_M)\\\nonumber
& &-\frac{1}{4}\left[\left\lbrace  3 + 5\left( \frac{\beta}{\alpha}\right) ^2 \right\rbrace  - \left\lbrace  3 +\left(  \frac{\beta}{\alpha} \right) ^2\right\rbrace  \cos^2(\sqrt{3}\beta h_M)\right] =0.
\end{eqnarray}
%%%%%%%%%%%%%%%
In \figurename{ \ref{30evenModes}} we show the symmetric longitudinal modes derived
with the values of $\alpha$ and $\beta$ obtained
from relations \eqref{alfabetakappa} and \ref{relalfabetakappa}
for $\kappa=2$, see also Table 2.

By \eqref{Ebc}
the free surface corresponding to the above symmetric and antisymmetric
modes is computed by 
\begin{equation*}
\eta(z)= \frac{1}{g}\frac{\partial \phi}{\partial y}\mid _{y=h_M}.
\end{equation*}%\label{freeS}
\begin{figure}%[h!]
\captionsetup{width=0.83\textwidth}
\centering
		\includegraphics[scale=0.36]{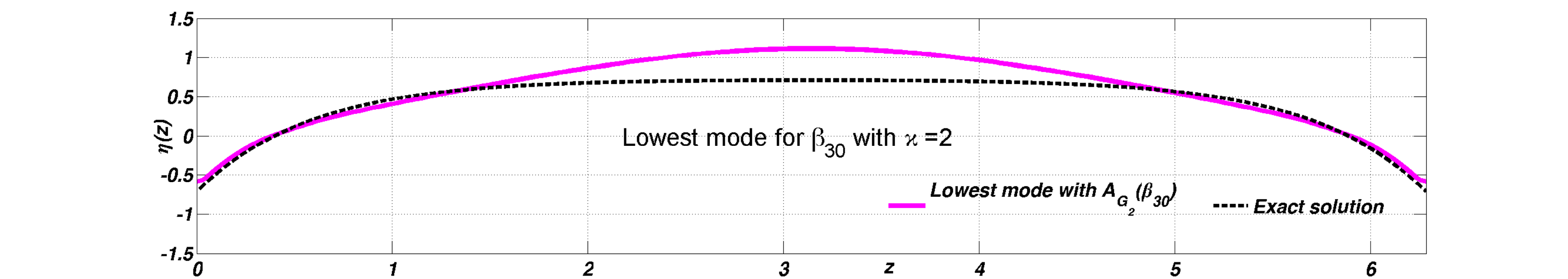}
\caption{Lowest longitudinal mode with  $\kappa=2$ given by \protect\eqref{lowestmode} and  \protect\eqref{lowestmodefreq}
for the triangular channel illustrated in \figurename{ \protect\ref{Figtrian30}}.} \label{30lowestMode}
\end{figure}
\begin{figure}
\captionsetup{width=0.83\textwidth}
\centering
		\includegraphics[scale=0.36]{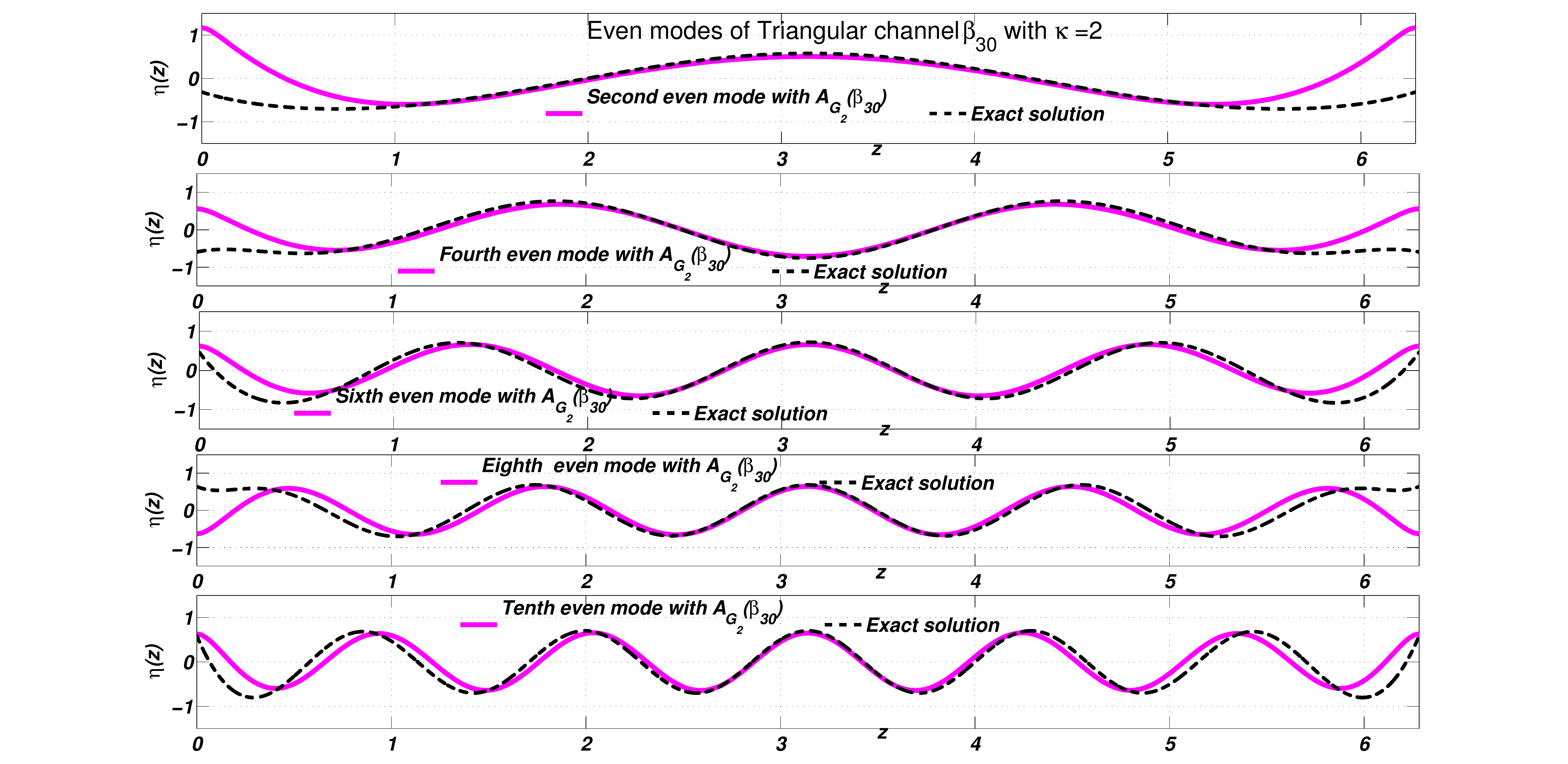}
\caption{
%The symmetric modes $2$, $4$, $6$, $8$, $10$ given by \eqref{sym30} and \eqref{omegasym30} for the triangular channel given by \figurename{ \ref{Figtrian30}} with analytic expression given by \eqref{beta30}. With $\kappa=2.$}
Symmetric longitudinal modes: $2$, $4$, $6$, $8$, $10$  of a channel with triangular cross 
section illustrated in \figurename{ \protect\ref{Figtrian30}}.
Magenta (solid) lines: operator $\mathcal{A}_{G_{\kappa}}(\beta_{30})$ with $\kappa=2$, dashed lines: 
exact solutions  given by \protect\eqref{sym30} and \protect\eqref{omegasym30} and the values in Table 2.}
\label{30evenModes}
\end{figure}
\begin{figure}%[h!]
\captionsetup{width=0.71\textwidth}
\centering
		\includegraphics[scale=0.37]{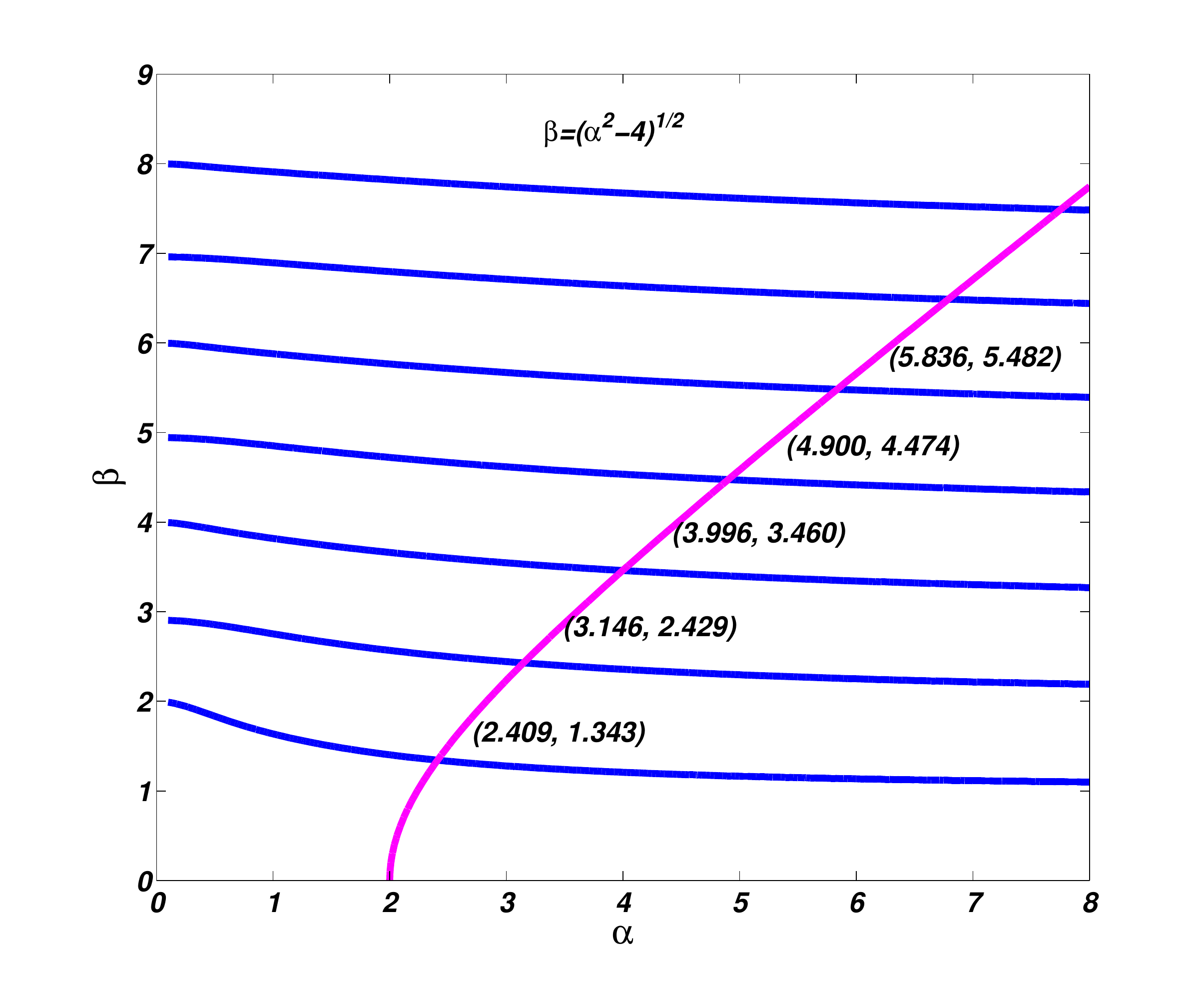}
\caption{Graphs for the determination of $\alpha$ and $\beta$ for the
modes $2$, $4$, $6$, $8$, and $10$ using $\kappa=2$,
$h_M=\frac{\pi}{\sqrt{3}}$.} \label{30locus}
\end{figure}
\begin{table}%[h!]
  \centering
  \captionsetup{width=0.71\textwidth}
  \caption{Frequencies of modes of channel of \figurename{ \ref{Figtrian30}}}
  \begin{tabular}{ccccccc}
   \toprule
$\text{ } $ &$i=2$ & $i=4$& $i=6$ & $i=8$ & $i=10$ & $\cdots$ \\
   \midrule
    $\alpha_i$ & 2.409 & 3.146 & 3.996 & 4.900 & 5.836 & $\cdots$\\
    $\beta_i$ & 1.343 & 2.429 & 3.460 & 4.474 & 5.482 & $\cdots$\\
    $\omega_i$ &2.4297   & 2.7764  &3.1289 & 3.4651  &3.7813  & $\cdots$\\
  \bottomrule
  \end{tabular}
  \caption*{\small{These values are associated to graphic roots in \figurename{ \ref{30locus}}
  for the geometry with the cross-section of \figurename{ \protect\ref{Figtrian30}}, 
  see \protect\eqref{beta30}. We use $h_M=\frac{\pi}{\sqrt{3}}.$}}
\end{table}\label{tab:table3p1}
In \figurename{s \protect\ref{30lowestMode}}
and  \protect\ref{30evenModes}
we compare the surface amplitude of 
the exact symmetric modes  with  the 
surface amplitudes obtained by computing numerically the eigenfunctions of
the approximate Dirichlet-Neumann operator 
$ {\mathcal{A}}_{G_{\kappa}}(\beta)$ of \eqref{Aproxgorro-3D} with 
$\beta$ as in \eqref{beta30}.
We use  $\kappa=2$. To compute the 
eigenfunctions of ${\mathcal{A}}_{G_{\kappa}}(\beta)$ numerically we  
use $2 \pi-$periodic boundary conditions.
Also, given a computed eigenfunction $f$ the surface amplitude $\eta$
is given by $\eta = \frac{1}{g}\mathcal{A}_{G_{k}}(\beta)$.  This is analogous to \eqref{freeS}. 

\figurename{ \ref{30evenModes}}  shows
good quantitative agreement for the even modes in the interior,
with some discrepancies at the boundary that represents the sloping beach.

\begin{figure}%[h!]
\captionsetup{width=0.71\textwidth}
\centering
		\includegraphics[scale=0.52]{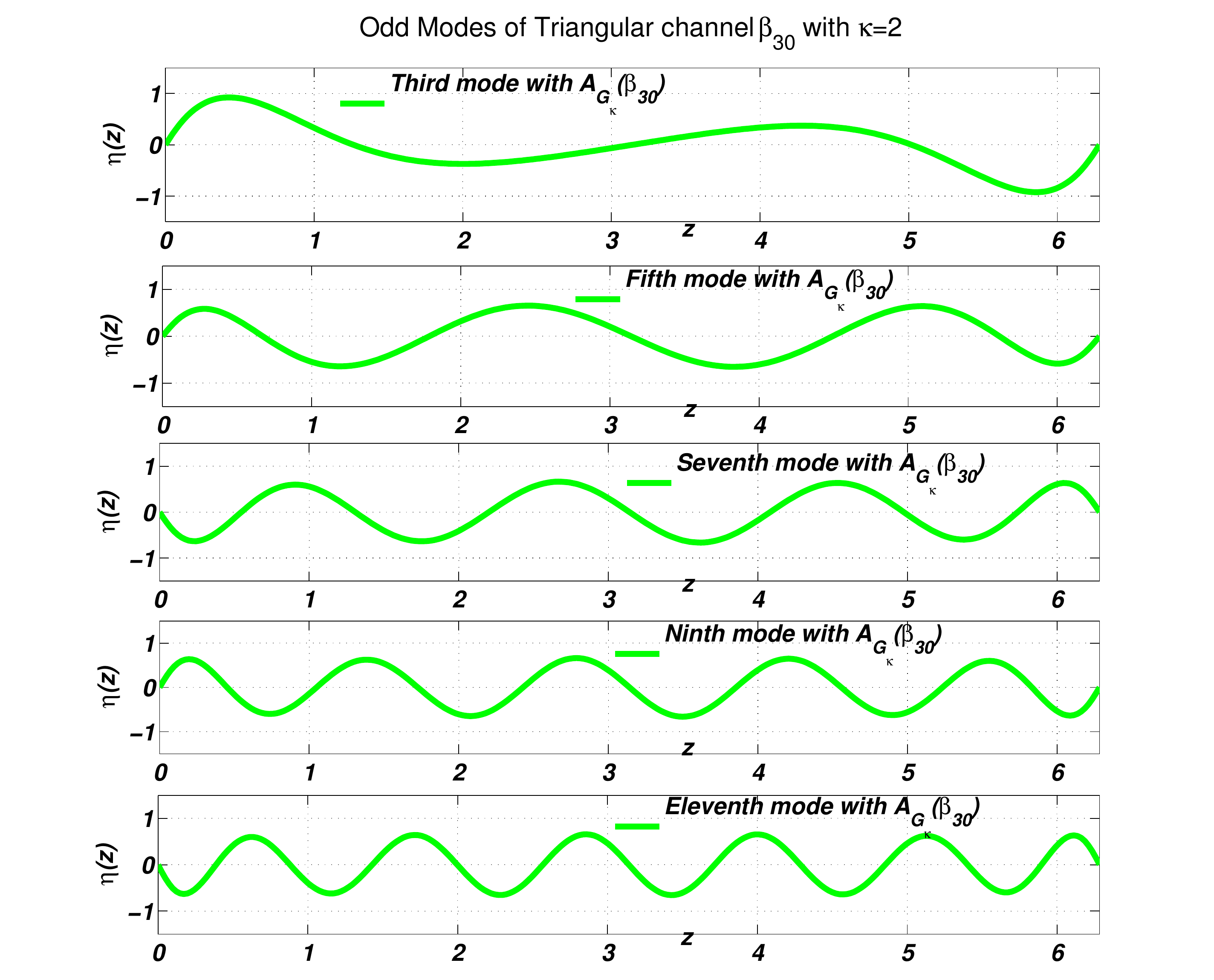}
\caption{Odd longitudinal modes: $3$, $5$, $7$, $9$  derived from operator $\mathcal{A}_{G_{\kappa}}(\beta_{\beta_{30}})$ with $\kappa=2$ for straight channel with triangular cross-section illustrated in \figurename{ \ref{Figtrian30}}.}
%Green (solid) lines:  operator $\mathcal{A}_{G_{\kappa}}$.}
\label{oddmodesTriangle30}
\end{figure}

To our knowledge there are no exact solutions reported in the literature for odd modes. 
Odd modes obtained 
with the approximate Dirichlet-Neumann operator are 
shown in \figurename{ \ref{oddmodesTriangle30}}.
The results on  the isosceles triangle of angle $45^\circ$ suggest 
that the solutions should be accurate far from the boundary.

\begin{figure}%[h!]
\captionsetup{width=0.7\textwidth}
\centering
		\includegraphics[scale=0.33]{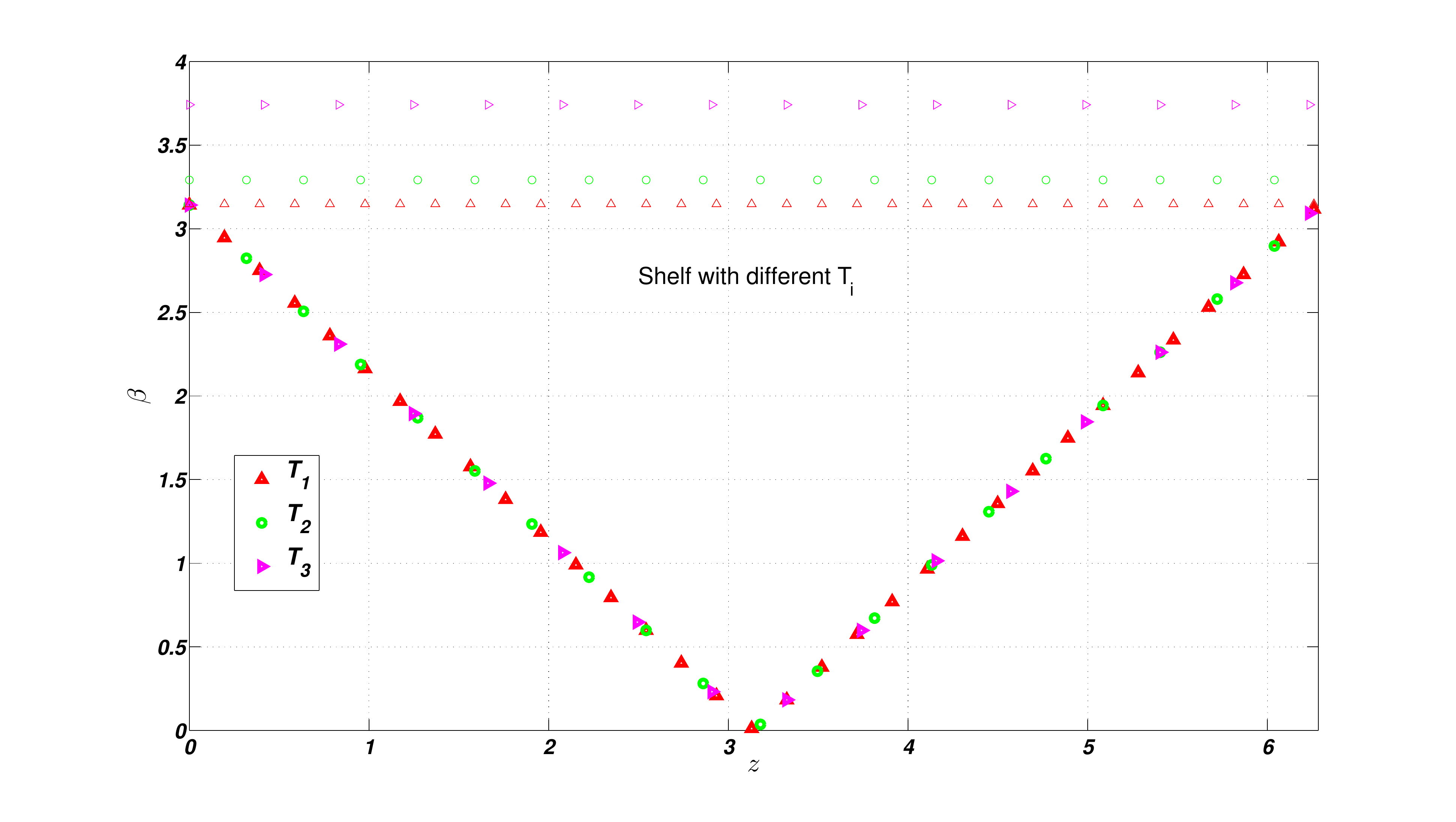}
\caption{Cross section of the straight channels with 
lateral boundary $\Gamma_L$. Vertical segments of height $T_1=0.006$ (red), $ T_2=0.15$ (green), $T_3=.6$ (magenta).} \label{CSt1t2t3}
\end{figure}

\begin{figure}%[h!]
\captionsetup{width=0.85\textwidth}
\centering
		\includegraphics[scale=0.4]{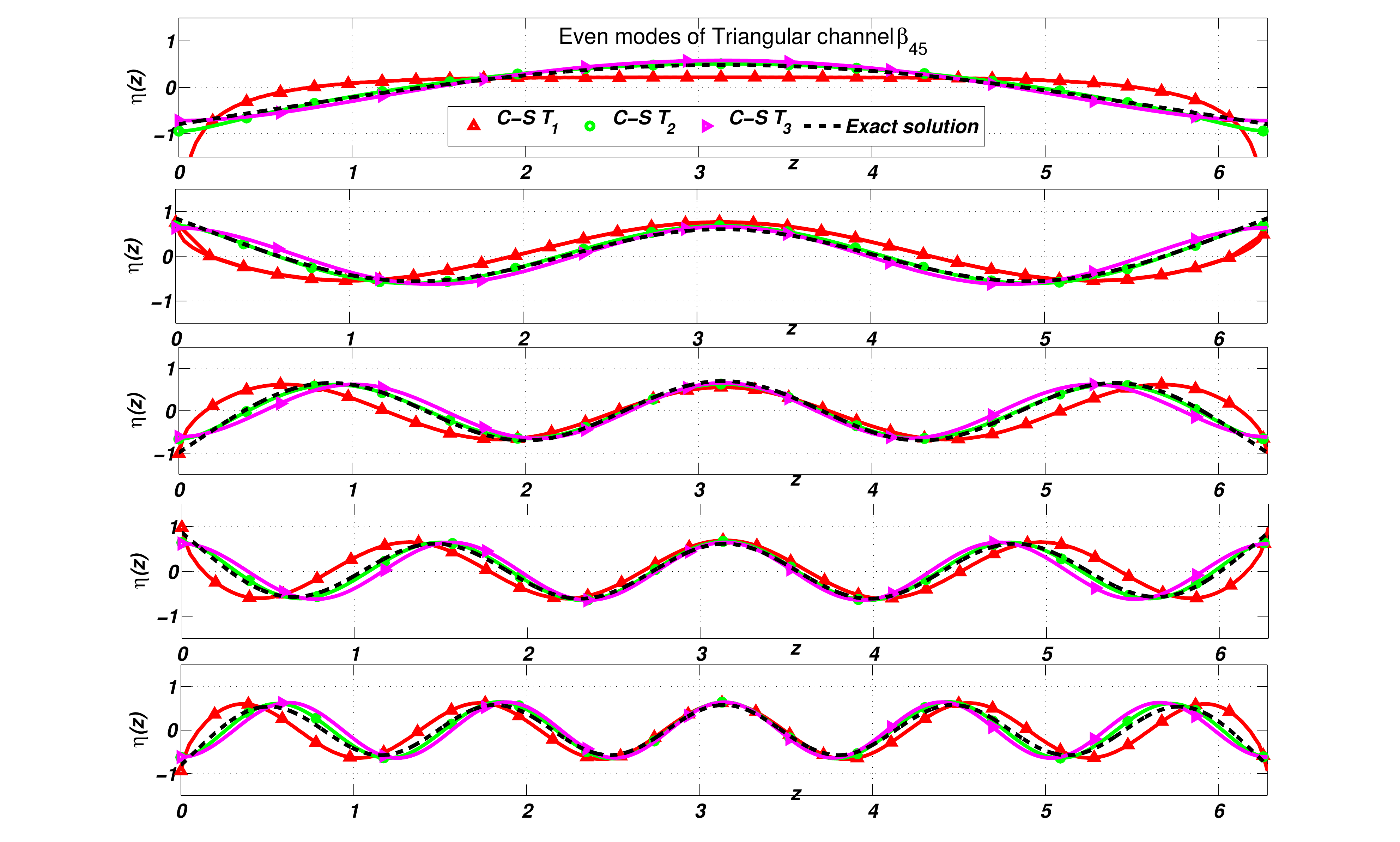}
\caption{Even modes for the three cross sections of \figurename{ \protect\ref{CSt1t2t3}}, vertical segments of
heights $T_1=.006$ (red), $ T_2=.15$ (green) , $T_3=.6$ (magenta).} \label{EvenmodesCSt1t2t3}
\end{figure}

\begin{figure}%[h!]
\captionsetup{width=0.85\textwidth}
\centering
		\includegraphics[scale=0.4]{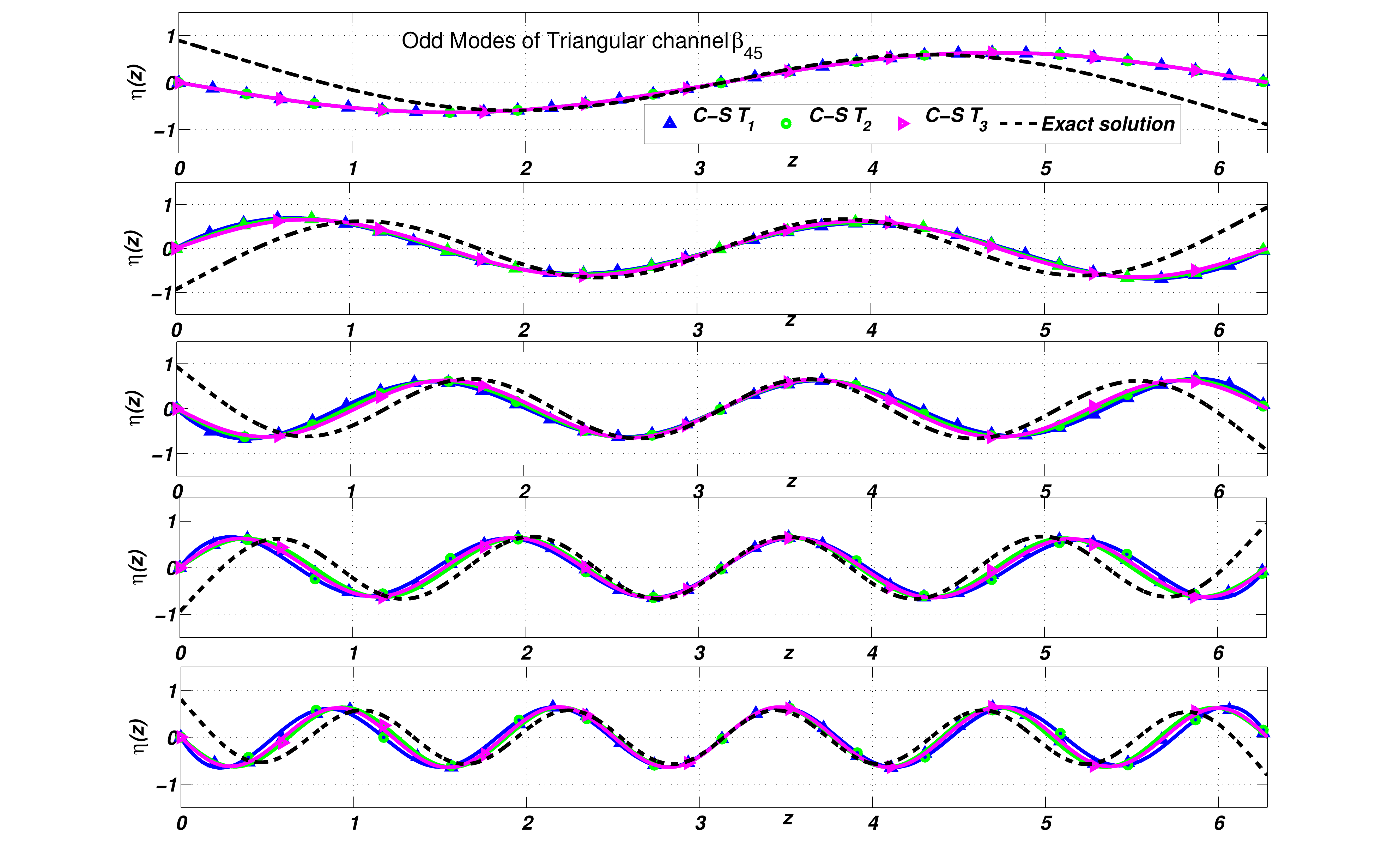}
\caption{Odd modes 
 for the three cross sections of 
\figurename{ \protect\ref{CSt1t2t3}}
vertical segments of
heights $T_1=.006$ (red), $ T_2=.15$ (green) , $T_3=.6$ (magenta).}
\label{OddmodesCSt1t2t3}
\end{figure}

We have also used the approximate Dirichlet-Neumann operator
to compute the $2 \pi-$periodic 
normal modes of domains obtained from the triangular channel 
by adding an interval of extra depth, see \figurename{ \ref{CSt1t2t3}}. 
We denote the added depth by $T$.  
\figurename{ \ref{EvenmodesCSt1t2t3}}, and   
\figurename{ \ref{OddmodesCSt1t2t3}} 
indicate the convergence to the triangular domain modes as $T$
vanishes. In this geometry 
even modes satisfy a Neumann boundary condition at $z = 0$, $2 \pi$, and  
$T$ can be interpreted also as the height of a vertical wall. 

\section{Trapped modes over continental shelf profiles}

\begin{figure}
    \centering
    \begin{subfigure}[b]{0.48\linewidth}        %% or \columnwidth
        \centering
        \includegraphics[scale=0.34]{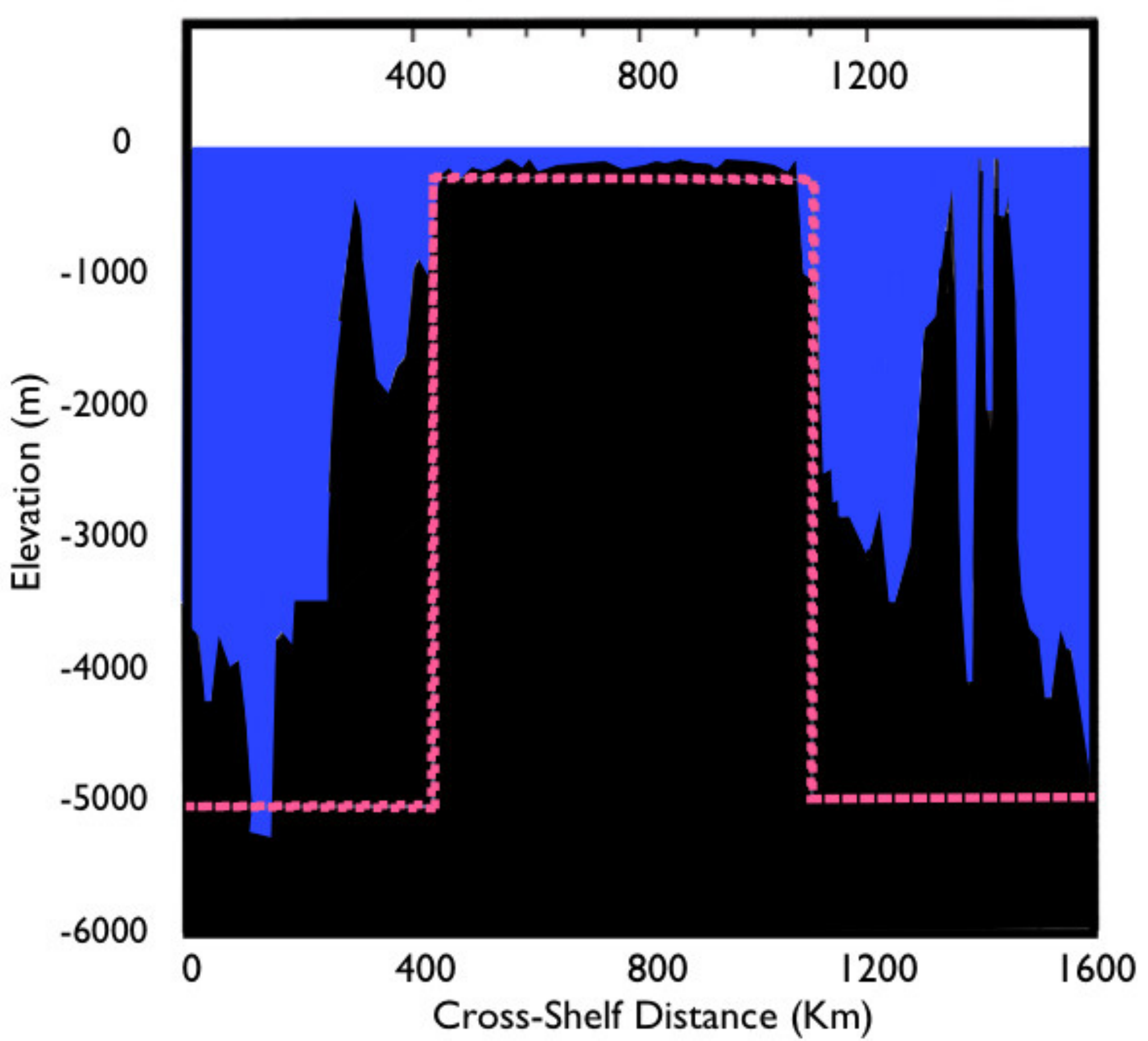}
        \caption{Off coast of Taiwan.}
        \label{fig: FigTaiwanblue}
    \end{subfigure}
    \begin{subfigure}[b]{0.48\linewidth}        %% or \columnwidth
        \centering
        \includegraphics[scale=0.27]{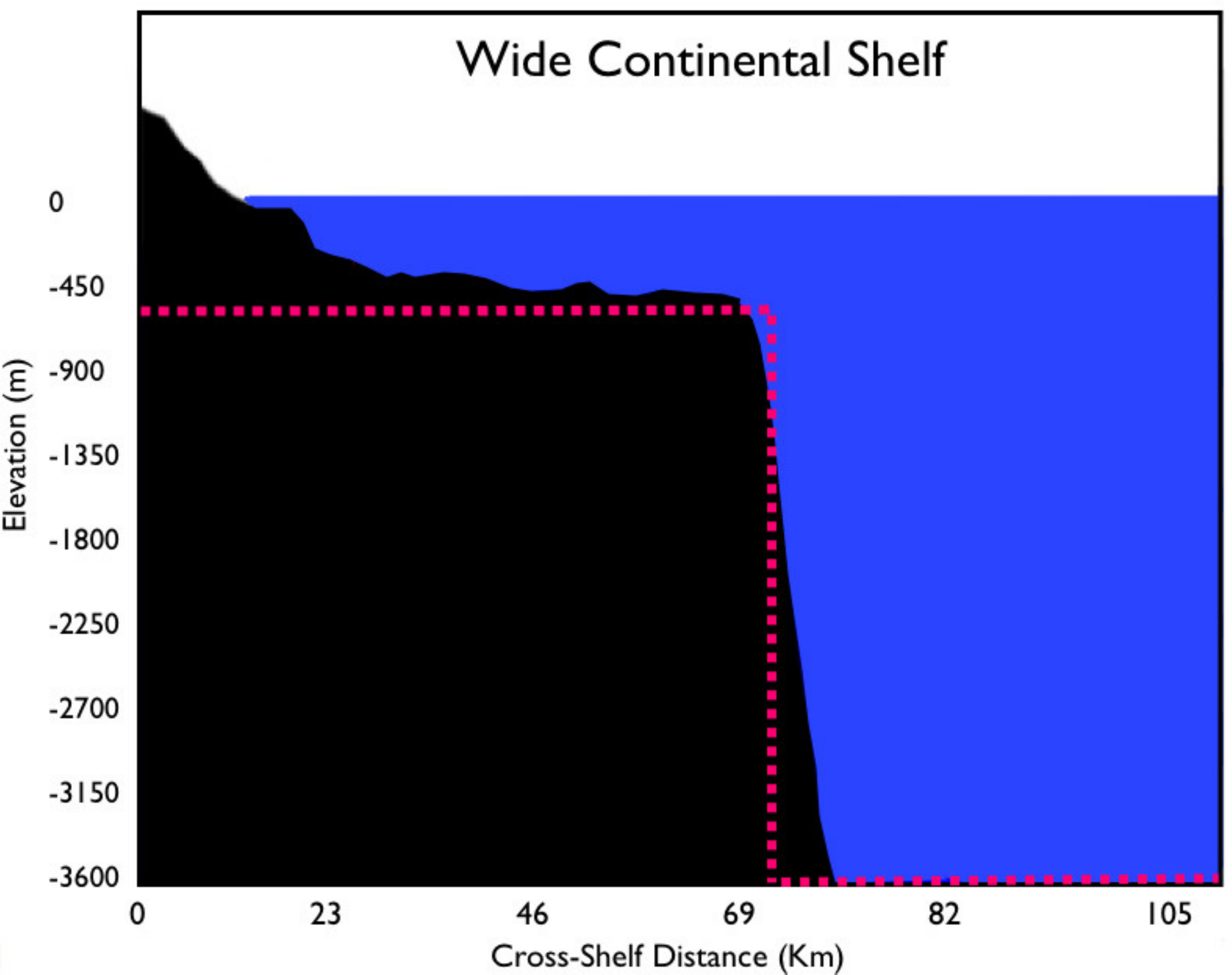}
        \caption{Coast of California.}
        \label{fig:FigOffCalifblue}
    \end{subfigure}
    \caption{a)  Typical topographic profile of shelf at distance of $50 km$ from the
    coast of Taiwan, \protect\cite{lin2001anomalous} . b) Typical topographic profile of  California coast, \protect\cite{mei2005theory, miles1972wave}. }
    \label{fig:roc_curve}
\end{figure}

In this section we study a model problem of 
waves that travel along a continental shelf and decay in the transverse direction.  
These are examples of longitudinal modes \eqref{LongMode} and we
use a two-valued piecewise constant
transverse depth profile $\Omega_U$,
see \eqref{omega-unbounded}. 
We will compare results from two approximations of the linear water
wave system \eqref{Ebc}. The first is a variable depth version
of the linear shallow water equation, where the
spectral problem for piecewise constant depth
is reduced to solving algebraic equations,
see Miles \cite{miles1972wave}, Lin, Juang and Tsay \cite{lin2001anomalous}, and
\cite{mei2005theory}.
These studies model continental shelves off the coasts of
California and Taiwan respectively, and we will use the same parameters. 
The second approximation is 
the linear Whitham-Boussinesq equation \cite{vargas2016whitham},
whose longitudinal modes lead to the spectral problem for the model 
Dirichlet-Neumann ${\mathcal{A}}_{G_{\kappa}}(\beta)$ of
\eqref{Aproxgorro}. The numerical eigenfunctions of this operator 
are compared with the trapped modes obtained with the shallow water theory.

The problem considered in the literature \cite{miles1972wave,lin2001anomalous,mei2005theory} 
for the shallow water model is 
not the usual spectral problem of fixing $\kappa$ and finding 
$\omega$ in \eqref{LongMode}, rather the authors fix $\omega$
(determined by observations) and look for values of $\kappa$ leading to  
solutions that decay in the transverse direction (trapped modes). 
We compare these modes to numerical eigenfunctions 
of ${\mathcal{A}}_{G_{\kappa}}(\beta)$ with the $\kappa$ obtained from the shallow
water problem. 

The continental shelves will be modeled by domains of the form $\Omega_U$, 
see \eqref{omega-unbounded}, with a depth that is constant in the longitudinal 
direction $x$ and only depends on the transverse direction $z$. 
The simplest model of a continental shelf is a plateau of constant depth 
$h_2$ in an interval of length $2a$,  
and $h_1 > h_2$ for all other $z$. 
e.g. $h_M = - h_1 $, $h_m = 0$ in the notation of \textit{Section} \ref{formulation}.

\subsection[Exact trapped modes solutions with shallow water theory]
{Exact trapped modes solutions over continental shelves with shallow water theory.}\label{exatTrappeM}

We outline the linear shallow water theory of 
\cite{miles1972wave}, \cite{mei2005theory}, 
and \cite{lin2001anomalous}. The linear shallow water wave equation is  
\begin{equation}\label{prewave}
g \nabla \cdot  (h(z) \nabla \eta)= \frac{\partial^2 \eta}{\partial t^2}, 
\end{equation}
with $\nabla = (\partial_x,\partial_z)$. The function $h$ is the depth.
We are looking for longitudinal wave solutions of the form
\begin{equation}\label{CosMei}
\eta(x,z,t)=V(z) \cos(\kappa x -\omega t),
\end{equation}
see \eqref{LongMode}.
By \eqref{CosMei}, \eqref{prewave} we have 
\begin{equation}\label{withcos2}
- \frac{\partial }{\partial z} \left(  h(z) \frac{\partial V(z)}{\partial z}\right)  + 
h(z) \kappa ^2  V(z)= \frac{\omega^2}{g} V(z),  
\end{equation}

Assuming  $h(z)$ piecewise constant with 
\begin{equation}\label{hpiecewise}
  h(z)=\left\lbrace 
\begin{array}{l}   
h_2  \text{ for } z\in[-a ,a],\\
h_1 \text{ for } z \notin[-a,a],\\
\end{array} 
\right.  
\end{equation}
\eqref{withcos2} becomes  
\begin{equation}\label{withcos}
gh_j \frac{\partial^2 V}{\partial z^2}  +V\omega^2 =  \kappa ^2 gh_j V,  
\end{equation}
with $j = 1, 2$. We fix $\omega $ and 
solve the equation in each region.  
At each constant depth we have
\begin{equation}\label{ecSch}
 \frac{\partial^2 V}{\partial z^2} = (\kappa ^2-\lambda_j^2) V,
\quad\hbox{with}\quad  \lambda_j=\frac{\omega}{\sqrt{gh_j}},   
\end{equation}
with $j = 1, 2$. Clearly, we have two kinds of solutions:
\begin{itemize}
\item[1.]   \textit{Oscillatory solutions.} If $ \lambda^2  \geq \kappa^2$, then  
\begin{equation}
V(z)= Be^{i \alpha z} + Ce^{-i\alpha z}, \quad\text{where}\quad \alpha= \sqrt{\lambda^2-\kappa ^2}.
\end{equation}
\item[2.]\textit{Exponentially growing/decaying solutions.}  If $ \lambda^2  \leq \kappa^2$, then 

\begin{equation}
V(z)= Ae^{\gamma z} + De^{-\gamma z}, \quad\text{where}\quad \gamma= \sqrt{\kappa^2-\lambda^2 }.
\end{equation}
\end{itemize}

We seek solutions that are oscillatory for $z \in [-a, a]$,    
decay exponentially for $z \notin [-a, a]$, and lead to continuous  
$\eta$ and $h(z)\frac{\partial \eta}{\partial z}$. 
The continuity condition requires that 
that $V$, and $h(z) V$ be continuous at $z = a$, $-a$.
Note that the continuity condition 
guaranties that the solution is a weakly differentiable function.

By the symmetry of the equation and the domain it is enough to
look for even and odd real solutions.
Even solutions are given by 
\begin{equation}\label{fevenaa}
V(z)=B \cos(\alpha_2 z), \text{ } \text{ } z\in[-a,a],
\end{equation}
with 
\begin{equation}\label{alpha2}
\alpha_2 = \sqrt{\lambda_2^2-\kappa^2}, \quad \lambda_2= \frac{\omega}{\sqrt{gh_2}},
\end{equation}
and 
\begin{equation}\label{fevenfueraaa}
V(z)= A e^{-\gamma_1(\mid z \mid-a)},  \text{ } \text{ } z\notin[-a,a],
\end{equation}
with
\begin{equation}\label{gamma1}
\gamma_1= \sqrt{\kappa^2-\lambda_1^2}, \quad \lambda_1= \frac{\omega}{\sqrt{gh_1}}.
\end{equation}
 
Odd solutions are given by  
\begin{equation}\label{foddaa}
V(z)=B' \sin(\alpha_2 z) \text{ } \text{ at } z\in[-a,a] \text{ and } h(z)=h_2,
\end{equation}
with $\alpha_2$, $\lambda_2$ as in \eqref{alpha2}, 
and 
\begin{equation}\label{foddfueraaa1}
V(z)= -A' e^{-\gamma_1(\mid z \mid-a)}  \text{ } \text{ at } z\notin[-a,0] \text{ and } h(z)=h_1, 
\end{equation}
\begin{equation}\label{foddfueraaa2}
V(z)= A' e^{-\gamma_1(\mid z \mid-a)}  \text{ } \text{ at } z\notin[0,a] \text{ and } h(z)=h_1, 
\end{equation}
with $\gamma_1$, $\lambda_1$ as in \eqref{gamma1}.

Requiring continuity of $\eta$ and $h(z)\frac{\partial \eta}{\partial z}$ 
at $z= \pm a$ we have the following. 

For the even modes, \eqref{fevenaa}-\eqref{gamma1} imply 
\begin{equation}
B\cos(\alpha_2 a)=A,
\end{equation}

\begin{equation}
-h_2B\alpha_2\sin(\alpha_2 a)=-h_1\gamma_1A,
\end{equation}
which imply
\begin{equation}\label{tan}
\tan(\alpha_2 a)=\frac{h_1\gamma_1}{h_2\alpha_2}.
\end{equation}
We thus obtain an equation for 
$\alpha_2$ and $\gamma_1$. This equation can not be solved analytically. 
We thus search for a solution graphically (i.e. numerically).
To do so, we first let 
\begin{equation}\label{eqforalpha2}
\chi= a\sqrt{\lambda_2^2-\kappa^2}=a\alpha_2, \quad h=  \frac{h_2}{h_1}.
\end{equation}
Then we can rewrite equation \eqref{tan} as
\begin{equation*}\label{taneven}
\tan(\chi)=\frac{\sqrt{\kappa^2-\lambda_1^2}}{h\alpha_2}.
\end{equation*}
Note that $\lambda_1^2 = h\lambda_2^2$, and that
\begin{eqnarray}
a\sqrt{\kappa^2 - h \lambda_2^2}=\sqrt{a^2\kappa^2 - a^2h \lambda_2^2}
&=&\sqrt{
 \lambda_2^2 a^2(1- h) -
\chi^2 }, 
\end{eqnarray}
so that letting 
\begin{equation}\label{omega}
\chi_{*}^2=% \lambda_2^2 a^2 (1- h)=
\frac{(\omega a)^2}{gh_2}(1-h),
\end{equation}
we must solve
\begin{equation}\label{cotchi}
\cot \chi =\frac{h\chi}{\sqrt{\chi_*^{2}-\chi^2}}.
\end{equation}
Continuity of the odd solutions leads to 
\begin{equation}
B'\sin(\alpha_2 a)=A',
\end{equation}
\begin{equation}
h_2B'\alpha_2\cos(\alpha_2 a)=-h_1\gamma_1A’,
\end{equation}
implying
\begin{equation}\label{cotodd}
\cot(\alpha_2 a)=-\frac{h_1\gamma_1}{h_2\alpha_2}.
\end{equation}
Using the above notation this is equivalent to 
\begin{equation}\label{realodd}
\tan \chi=-\frac{h\chi}{\sqrt{\chi_*^{2}-\chi^2}}.
\end{equation}
The numerical solutions of
\eqref{cotchi}, \eqref{realodd}
are shown in \eqref{cotodd}, \eqref{realodd} and in Tables \ref{tab:table1} and \ref{tab:table3}. 

%In fact from equations \eqref{cotchi} and \eqref{realodd}  
%we have given $\omega$ two  transcendental equations 
%for $\kappa$ for even and odd solutions respectively. This point will be relevant in the following sections.

\subsection{Trapped modes on idealized shelves off California and Taiwan coasts}

\textit{ Shelf off coast of California.} 
The first example we consider models the continental shelf off the coast of California 
following \cite{miles1972wave}, \cite{mei2005theory}. 
The depth topography is described by 
$h_2 = 600$ m, $h_1=3600 $ m, and total shelf length $2a= 140$ km, 
\figurename{ \ref{fig:FigOffCalifblue}}.
We compute both even and odd trapped modes, but by the geometry of \figurename{ \ref{fig:FigOffCalifblue}} we
 only consider the even ones,  restricted to $z>0$, as physical.

\begin{figure}
\captionsetup{width=0.71\textwidth}\label{Figbeta30}	
\includegraphics[scale=0.34]{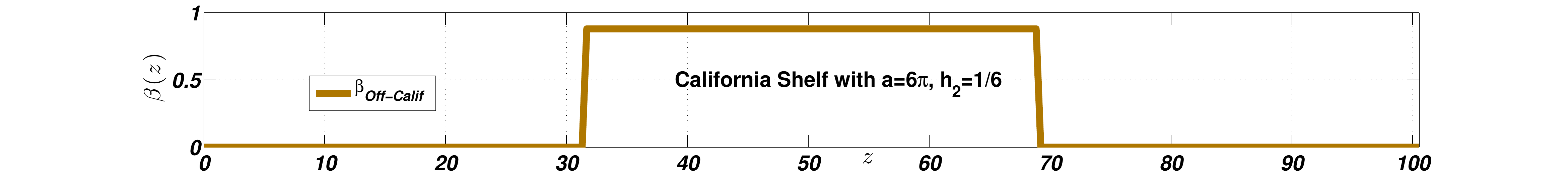}
%Fig11
\caption{California shelf, see \protect\eqref{betaOff}.} \label{FigOffC}	
\captionsetup{width=0.71\textwidth}
\end{figure}

In our computations we rescale,  
using $h_1=1$ so that $h_2= \frac{1}{6}=0.1617$. 
% and $h_M=1$ on both sides towards the sea. 
The rescaled total shelf length is $2a=12 \pi$, 
see \eqref{betaOff} for the corresponding 
cross-section in \figurename{ \ref{FigOffC}}. 
In the notation of \textit{Section} \ref{formulation},
$h_M =h_1 $, $h_m = 0$, and the bottom is at $y = h_m + \beta(z)$ with
\begin{equation}\label{betaOff} 
\beta(z)=\beta_{\text{Off-California}}(z) =\left\lbrace \begin{array}{l}   
0  \text{ in }  - \infty \leq z < 10\pi,\\
5/6  \text{ in } 10\pi \leq z \leq 22\pi,\\ 
0     \text{ in }  22\pi  \leq z < \infty.  \\ 
 \end{array} \right.
\end{equation}

The numerical eigenfunctions of the operator ${\mathcal{A}}_{G_{\kappa}}(\beta)$ of
\eqref{Aproxgorro} are computed using the domain $[0, 32 \pi]$, with periodic 
boundary conditions.  

\textit{Shelf off coast of Taiwan.} 
This continental shelf is located off the eastern coast of Taiwan, 
on the edge of the continental shelf of Asia. Off the edge of the shelf the 
slope plunges down to the deep Pacific Ocean, 
at a gradient of 1:10 and the ocean reaches a depth of more 
than 5000 meters about 50 kilometers off the coast,  
as shown in \figurename{ \ref{fig: FigTaiwanblue}}. 
An idealized model
for this shelf
by Lin, Juang and Tsay \cite{lin2001anomalous}
uses $h_2=80m$ on the shelf, $h_1=5000m$ 
on both sides of the ridge and total shelf length of $2a= 650km$,
see \figurename{ \ref{fig: FigTaiwanblue}}, \cite{lin2001anomalous}.

After rescaling, we consider a fluid domain  with depth $h_1=1$, 
obtaining $h_2= .0160$ over the ridge and  
a total shelf length is $2a= 130$,
see \eqref{betaT} for the corresponding  cross-section as shown \figurename{ \ref{taiwanshelf2}}.
In the notation of \textit{Section} \ref{formulation},
$h_M =h_1 $, $h_m = 0$, and the bottom is at $y = h_m + \beta(z)$ with
\begin{equation} \label{betaT}
\beta(z)=\beta_{\text{Off-Taiwan}}(z) =\left\lbrace \begin{array}{l}   
0  \text{ in }  - \infty \leq z < 25\pi, \\
24/25  \text{ in } 25\pi \leq z \leq 65\pi, \\ 
0     \text{ in }  65\pi \leq z < \infty. \\ 
 \end{array} \right.
\end{equation}

The numerical eigenfunctions of the operator ${\mathcal{A}}_{G_{\kappa}}(\beta)$ of
\eqref{Aproxgorro} are computed using the domain $[0, 90 \pi]$, with periodic 
boundary conditions.

\begin{figure}%[h!]
\captionsetup{width=0.65\textwidth}
\centering
		\includegraphics[scale=0.31]{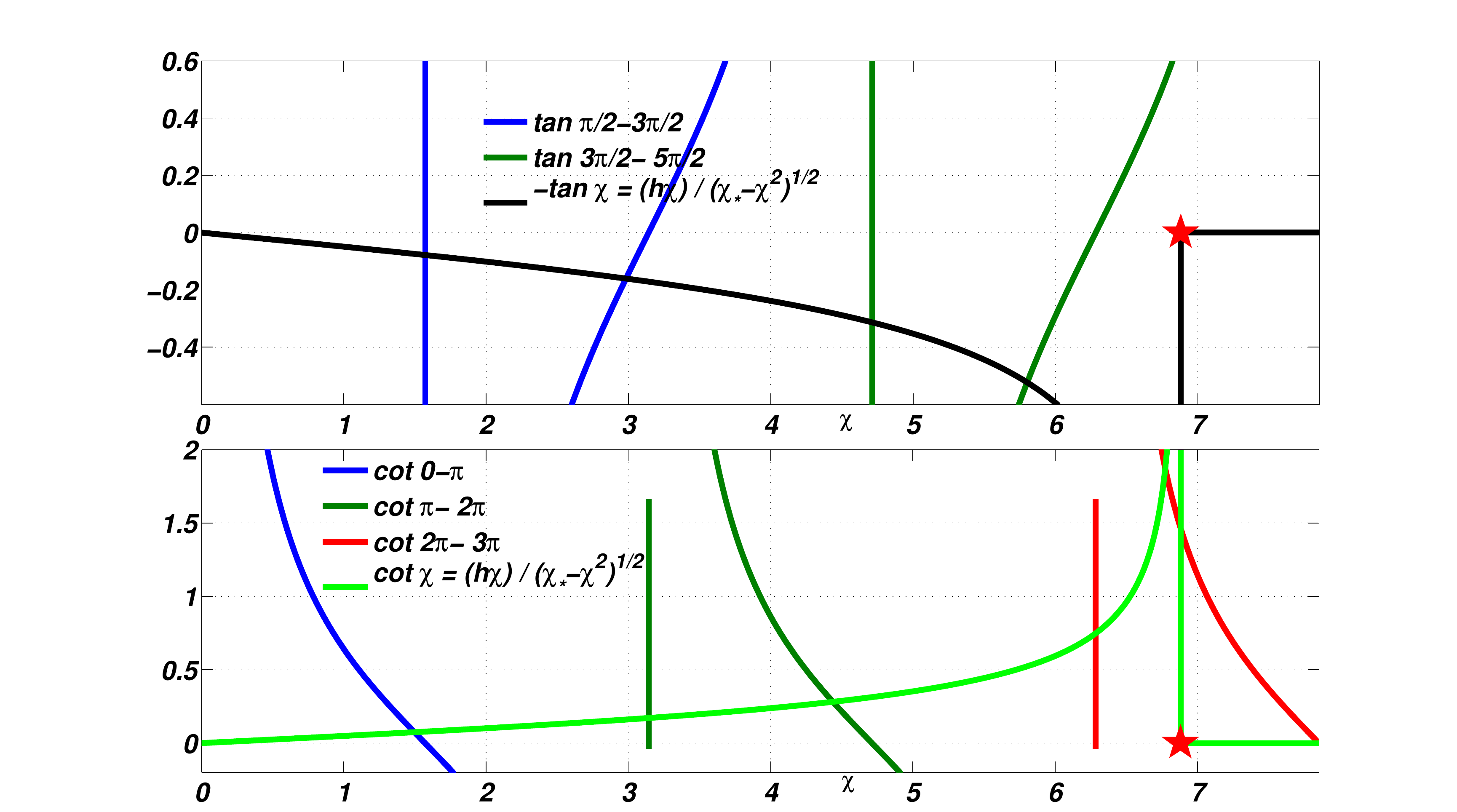}
\caption{Upper side shows graphical solution of  \eqref{realodd}  for odd modes of \textit{ Off Coast of California} and lower side shows graphical solution of \eqref{cotchi} for even modes. Red star corresponds to $\chi_{*}= 2.19\pi$} \label{CurveTannhcotOffCalif}	
\end{figure}

\begin{figure}%[h!]
\captionsetup{width=.8\textwidth}
\centering
\includegraphics[scale=0.33]{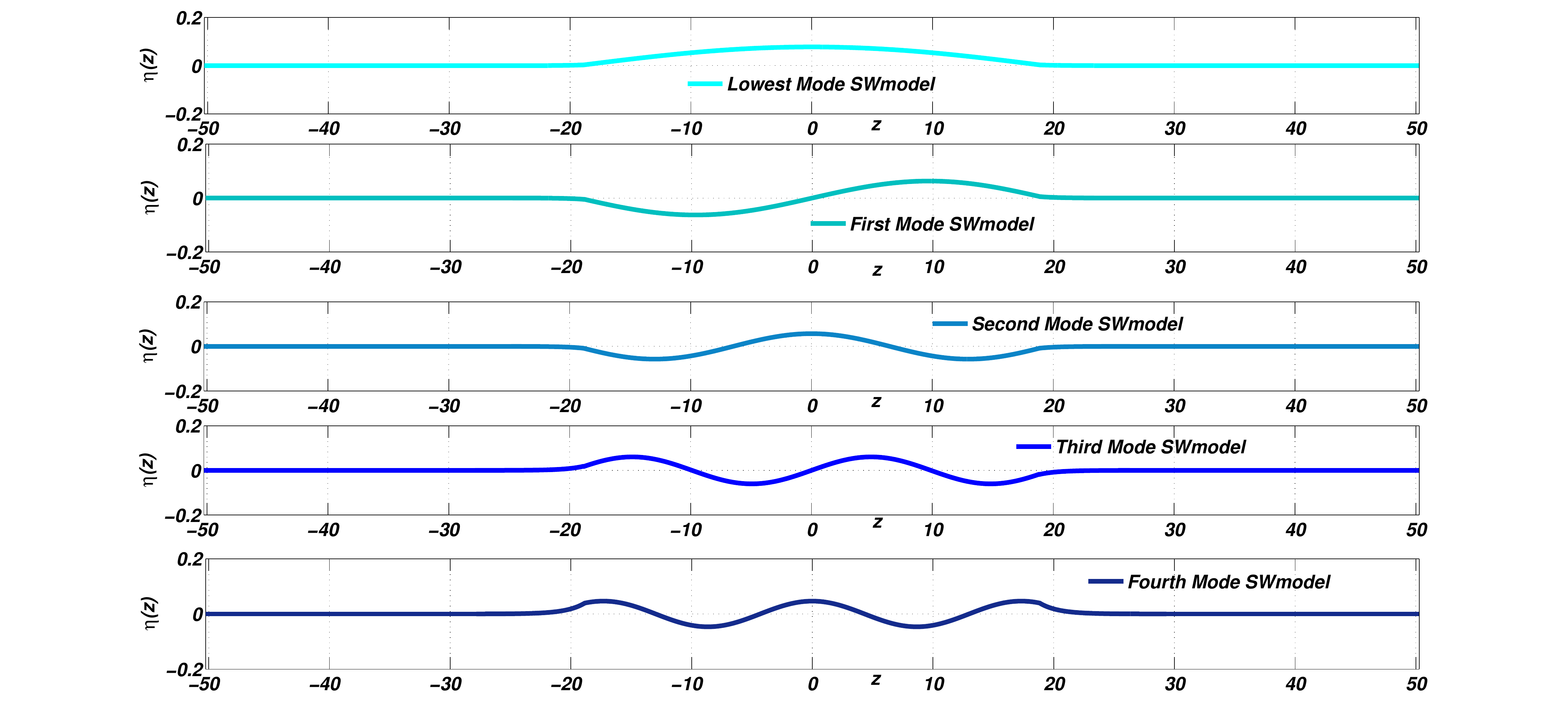}
%Fig15
\caption{Surface profiles of trapped modes derived with formulas \protect\eqref{fevenaa}
and \protect\eqref{foddaa} for the Coast of California, see 
\eqref{betaOff}, \figurename{ \protect\ref{FigOffC}}.} \label{plotMODESCalif}	
\end{figure}

\textbf{I. California shelf}
We use the data of \cite{miles1972wave} to compute
$\chi_*=6.8801\approx 2.19 \pi$, $\omega= 0.1582$.
From the graphical solutions shown
in the lower plot of \figurename{ \ref{CurveTannhcotOffCalif}} for $n$ even we see that for
$\chi_{*}= 2.19 \pi$
we have three points of intersection:
%between $y_1= \cot \chi $ and \eqref{cotchi};
$0 < \chi_0< \pi <\chi_2 < 2 \pi < \chi_4< \chi_* < 3\pi$.
From the graphical solution in the upper plot of the same figure
% \figurename{ \ref{CurveTannhcotOffCalif}}
for $n$ odd we see that for $\chi_{*}= 2.19 \pi$
there are two points of intersection:
 %between $y_3= \tan(\chi)$ and \eqref{realodd};
 $ \frac{\pi}{2} < \chi_1< \frac{3 \pi}{2}<\chi_3< \frac{5\pi}{2} $.
We thus have five trapped modes.
 In \tablename{ \ref {tab:table1}} we summarize the values $\chi_{i}$ obtained from
 the intersection of the curves.
In \tablename{ \ref {tab:table3}} we show
the values of  $\kappa_n$ and $\gamma_{1_n}$ for $n=0,\dots,4$.
The surface amplitude of each mode, see
\eqref{CosMei}, \eqref{fevenaa}-\eqref{foddfueraaa2},
is shown in \figurename{ \ref{plotMODESCalif}}.

\begin{table}%[h!]
  \centering
  \caption{The roots $\chi_n$ of \eqref{cotchi}, \eqref{realodd}}
  \label{tab:table1}
  \begin{tabular}{cccccc}
   \toprule
$\text{ } $ &$\chi_0$ & $\chi_1$& $\chi_2$ & $\chi_3$ & $\chi_4$\\
    \midrule
    Even & 1.5327 & \text{ }& 4.5656& \text{ } & 6.8429\\
    Odd & \text{ }  & 3.059 & \text{ } & 5.9954 & \text{ } \\
   \bottomrule
  \end{tabular}
%  \caption*{}
\end{table}
\begin{table}%[h!]
  \centering
  \caption{Wavenumbers $\alpha_n$, see \eqref{eqforalpha2}  associated to the roots $\chi_n$  }
  \label{tab:table3}
  \begin{tabular}{cccccc}
    \toprule
\text{ } & $n=0$ & $n=1$& $n=2$ & $n=3$ & $n=4$ \\
   \midrule
    $ \alpha_{2_n}= \frac{\chi_n}{a}$ &0.0813& 0.1623 &0.2422 & 0.3181& 0.3630 \\
     $\kappa_n=\sqrt{ \lambda_2^2-\alpha_{2_n}^2}$&0.6271& 0.6111 &0.5841 &0.5465  &0.5177\\
%    $ \delta_n $&0.1626 &0.3246 &0.4844&0.6361 & 0.7261 \\
    $ \gamma_{1_n}=\sqrt{ \kappa_n^2-\lambda_1^2}$&0.7460 &0.7326 &0.7102&0.6796 &0.6567 \\
   \bottomrule
  \end{tabular}
 \caption*{}
\end{table}

\begin{figure}%[h!]
\captionsetup{width=0.91\textwidth}
\centering
\includegraphics[scale=0.45]{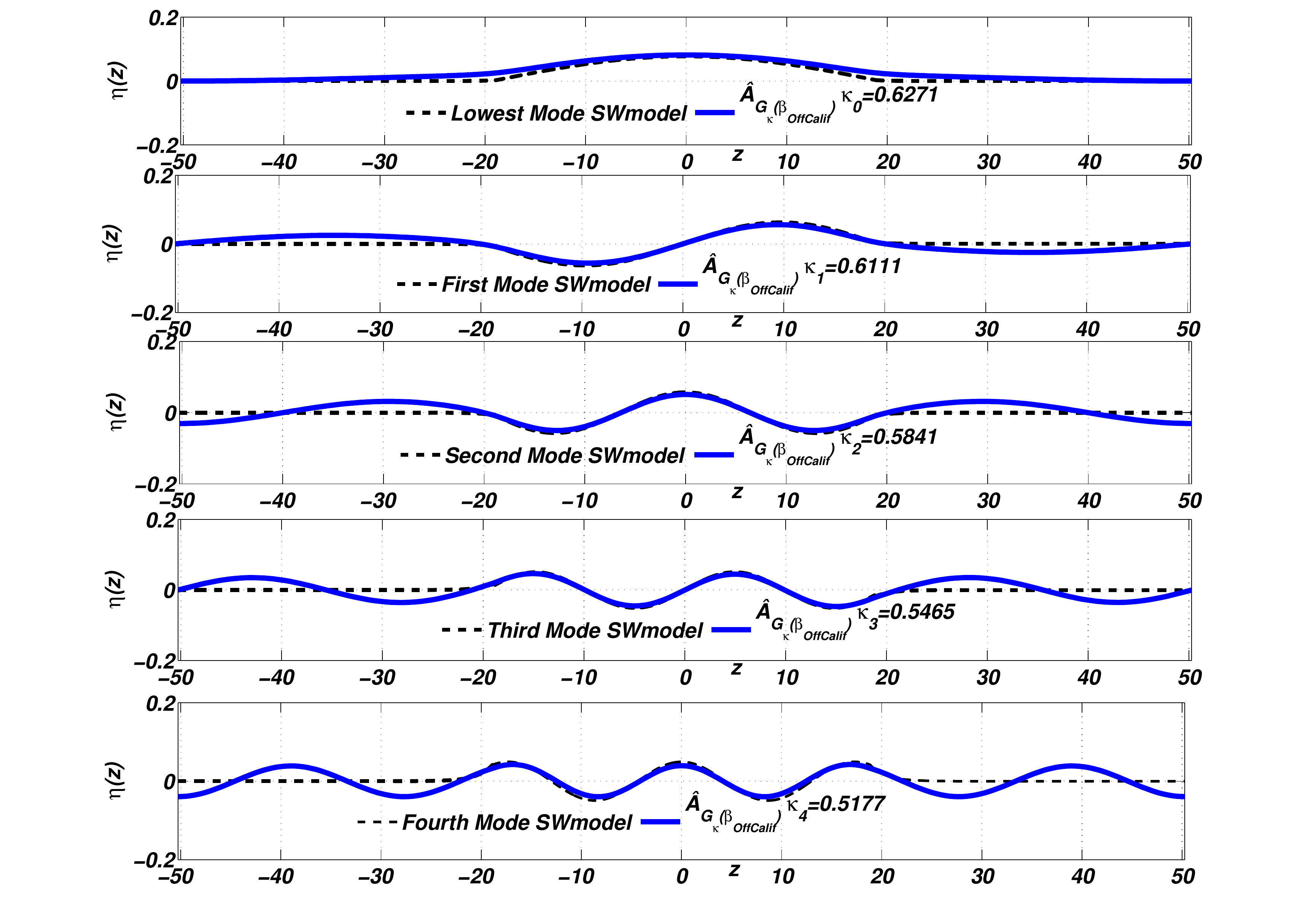}
%Fig15
\caption{ Comparisons of amplitude profiles of normal modes obtained by shallow water theory in  dashed lines and eigenfunctions of operator $\mathcal{A}_{G_{\kappa_n}}(\beta_{OffCalif})$ in solid lines.   
The figures from the top to the bottom shows:
in blue line  the surface profiles derived from operator$\mathcal{A}_{G_{\kappa_n}}(\beta_{OffCalif})$ for $n=0$, $\kappa_0=0.6271$, for $n=1$, $\kappa_1=0.6111$, $n=2$, $\kappa_2=0.5841$,
for $n=3$, $\kappa_3=0.5465$, $n=4$, $\kappa_4=0.5177$.
} \label{ModesSWvsOP}	
\end{figure}

To compare to the modes obtained using the operator 
${\mathcal{A}}_{G_\kappa}(\beta)$, we let $\kappa = \kappa_n$, with $n=0,\ldots,4$,
and for each value $\kappa = \kappa_n$
compute the corresponding eigenvalues and eigenvectors.
We compare the $n-$ th eigenvector of $\mathcal{A}_{G_{\kappa_n}}$ to 
the shallow water mode corresponding to
$\kappa_n$. The modes obtained are shown in 
\eqref{ModesSWvsOP}, together with the modes obtained using the shallow
water modes. We see that the two results are close, especially for the lowest modes. 

\begin{figure}%[h!]
\captionsetup{width=0.71\textwidth}
\centering
		\includegraphics[scale=0.28]{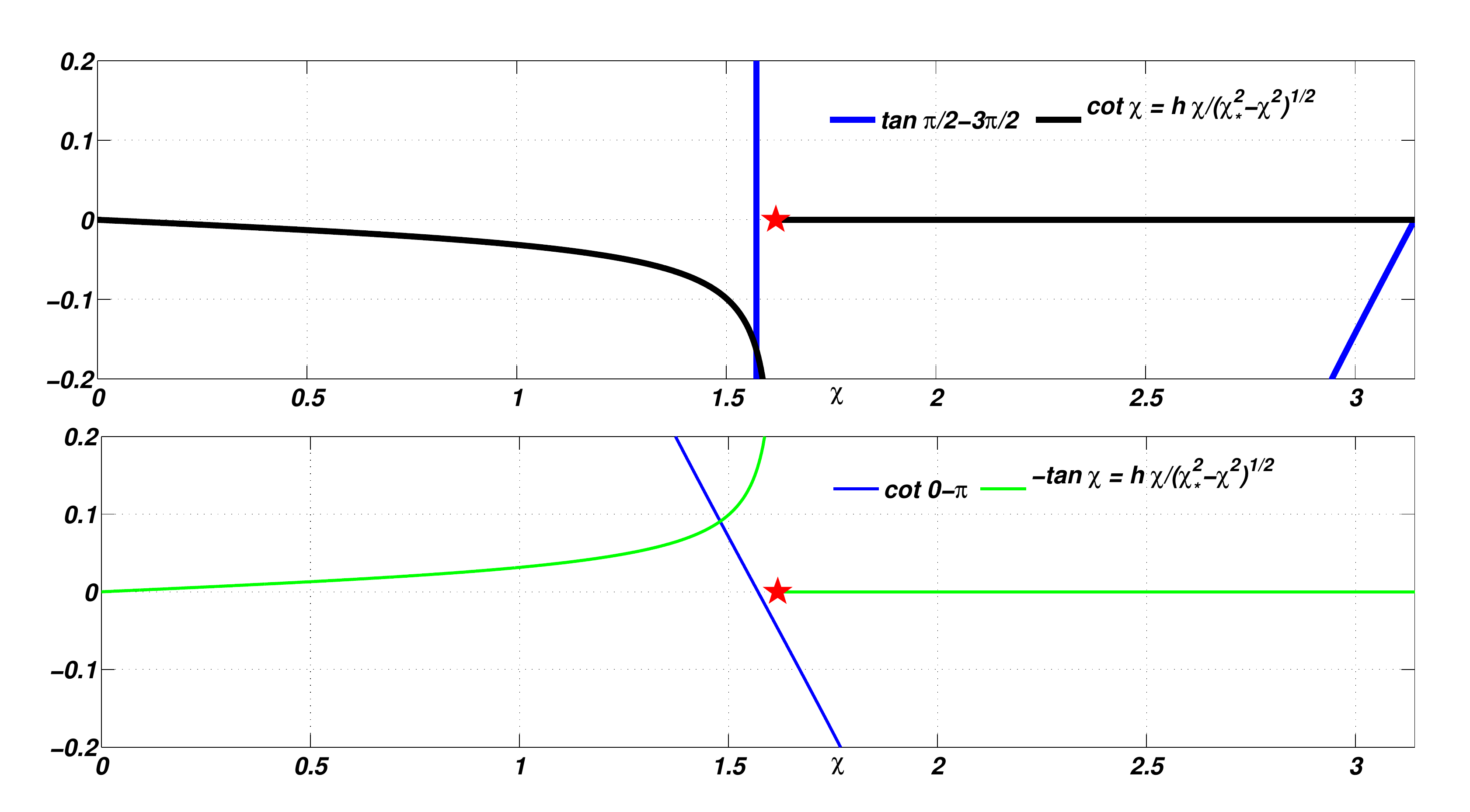}
\caption{Graphical solution of  \protect\eqref{realodd}  for odd modes of \textit{ Off Coast of  Taiwan} and lower side shows graphical solution of \eqref{cotchi} for even modes. Red star corresponds 
to $\chi_{*}= 1.618$.}	\label{fig18}
\end{figure}

\begin{figure}
\captionsetup{width=0.91\textwidth}
\centering
\includegraphics[scale=0.34]{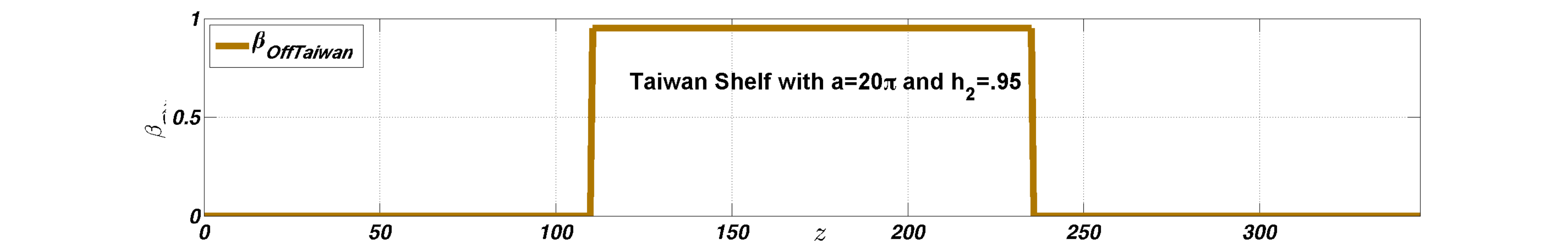}
\caption{Taiwan shelf, see \eqref{betaT}.}	
\captionsetup{width=0.84\textwidth}\label{taiwanshelf2}
\end{figure}

\begin{figure}\label{LowestTaiwan} 
\captionsetup{width=.9\textwidth}
\centering
\includegraphics[scale=0.35]{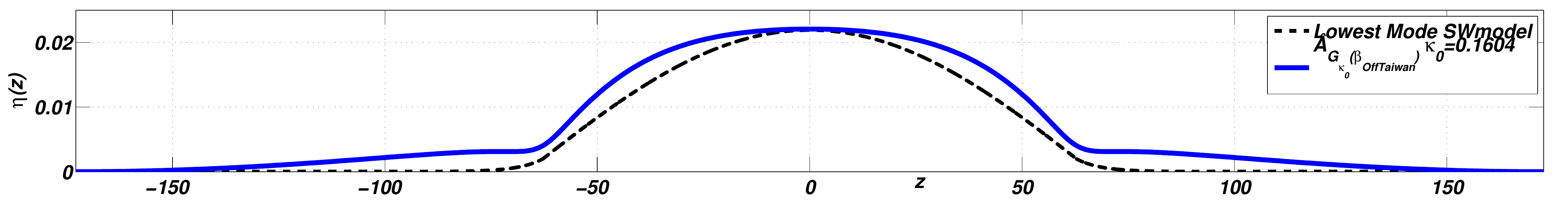}
\caption{Comparisons of amplitude profile of normal mode obtained by shallow water theory in dashed line and eigenfunction of operator $\mathcal{A}_{G_{\kappa_0}}(\beta_{OffTaiwan})$ for $n=0$, $\kappa_0=0.1604.$ in solid line.
 % In solid line is the surface profile derived from operator $\mathcal{A}_{G_{\kappa_n}}(\beta_{OffTaiwan})$ for $n=0$, $\kappa_0=0.1604.$
}\label{lowestmodetaiwan}	%\captionsetup{width=0.75\textwidth}
\end{figure}

\textbf{II. Taiwan shelf} 
In this example we follow 
\cite{lin2001anomalous},  obtaining $ \chi_{*}  = 1.618$, $\omega = 0.0165$. 
The graphical solution shown in \figurename{ \ref{fig18}} indicates that 
there is only one trapped even mode with $\chi_0= 1.4804$ and $\kappa = \kappa_0 = 0.1604$, see Table 5.
%\tablename{ \eqref{tab:table5}}. 
This mode is shown in in \figurename{ \ref{lowestmodetaiwan}}, where we also plot the lowest frequency
mode of ${\mathcal{A}}_{G_\kappa}(\beta)$. We observe that the two modes are close to each other.
\begin{table}%[h!]
  \centering
  \caption{Wavenumber $\alpha_0$, see \eqref{eqforalpha2} associated to the root $\chi_0$ }
  \begin{tabular}{cccccc}
  \toprule
\text{ } & $\alpha_{2_0}$ & $\kappa_0$& $\gamma_{1_0}$  \\
  \midrule
    $n=0$ & 0.0236&0.1604 &0.1760 \\
  \bottomrule
  \end{tabular}\label{tab:table5}
 \caption*{} 
\end{table}
\section{Discussion}
We have studied two problems on linear water wave modes in channels with
variable depth, choosing depth geometries and models with known exact results. 
The main goal was to test a nonlocal version of dispersive shallow water theory that 
uses a simple nonlocal approximation of the Dirichlet-Neumann operator for variable
depth. The approximate Dirichlet-Neumann operator we use 
leads to a unified and relatively simple way to compute numerically normal modes 
for variable depth channels.  
Our results on normal modes in bounded domains 
indicate that this operator can give reasonable approximations, 
even though  
geometries with exact normal modes have sloping beaches and seem to pose too stringent 
a test for the nonlocal operator. 
This is because our operator is most naturally defined on periodic 
functions and the symmetry of the depth profile 
imposes boundary conditions that are not present in the 
exact formulation of the normal mode problem. Despite this issue, 
the two approaches lead to normal modes that are close away from the boundary.
The main discrepancies are seen   
near the sloping beach for odd modes. These discrepancies 
motivate some further study 
on how to best use or extend 
the Dirichlet-Neumann approach for sloping beach problems. The operator used 
here was an ad-hoc approximation, but 
the boundary conditions problem seems relevant to more systematic 
approximations of the exact Dirichlet-Neumann operator, e.g. the ones in
\cite{craig2005hamiltonian,gouin2015development}.    

The problem of trapped modes for unbounded domains was studied using linear 
St. Venant shallow water theory. Despite the simplifications it introduces, shallow water   
theory has given some useful insights into the behaviour of trapped modes in other contexts, 
e.g. Ursell modes in semi-infinite sloping beach domains and their excitation by incoming
waves \cite{marin2007waves,romero2010trapped,bonnet1990mathematical,kuznetsov2001spectrum}.
Also, geophysical applications motivate simplified models that combines features 
of the both the St. Venant and variable depth 
Whitham-Boussinesq equations and it is therefore natural to compare the two models.  
However, the present study of the spectral 
problem for the transverse modes of the nonlocal linear equations  
is limited to reproducing results of the St. Venant theory, 
and is mainly indicative of the possible geophysical 
interest of the variable depth Whitham-Boussinesq equations.  
We are currently working on ways to determine the number of trapped modes of the operator 
$\mathcal{A}_{G_{\kappa}}$ for the continental shelf and similar depth topographies, 
some related results are in \cite{marin2007waves,romero2010trapped,kuznetsov2002linear}.   
It would be also of interest to consider nonlinear effects
on the modes discussed in this paper.

\section*{Acknowledgments}
We would like to thank especially Professor
Noel Smyth for many helpful comments.
Rosa Mar\'ia Vargas-Maga\~na was supported by Conacyt Ph.D. scholarship 213696. 
The authors also acknowledge partial support from grants SEP-Conacyt 177246 and PAPIIT IN103916.

\bibliographystyle{unsrt}

%\bibliography{WM2arxivbbl}

\end{document}